\documentclass[twoside,twocolumn,9pt]{article}
\usepackage{extsizes}
\usepackage[super,sort&compress,comma]{natbib} 
\usepackage[left=1.5cm, right=1.5cm, top=1.785cm, bottom=2.0cm]{geometry}
\usepackage{balance}
\usepackage[dvipsnames]{xcolor}
\usepackage{mathptmx}
\usepackage{sectsty}
\usepackage{graphicx} 
\usepackage{subfig}
\usepackage{pifont}
\usepackage{lastpage}
\usepackage[format=plain,justification=justified,singlelinecheck=false,font={stretch=1.125,small,sf},labelfont=bf,labelsep=space]{caption}
\usepackage{float}
\usepackage{fancyhdr}
\usepackage{fnpos}
\usepackage[english]{babel}
\addto{\captionsenglish}{%
  
}
\usepackage{array}
\usepackage{droidsans}
\usepackage{charter}
\usepackage[T1]{fontenc}
\usepackage{setspace}
\usepackage[compact]{titlesec}
\usepackage[hide links]{hyperref}

\usepackage{natbib}
\usepackage{amsmath,amssymb,mathrsfs,mathtools}
\usepackage{soul}
\usepackage{bbding}
\usepackage[normalem]{ulem}
\usepackage[version=4]{mhchem}

\usepackage{siunitx}

\definecolor{cream}{RGB}{222,217,201}
\newcommand{\pd}{{\mathsf{PD}}}
\newcommand{\tp}{{\mathsf{TP}}}
\newcommand{\vx}{\mbox{v}_x}

\def\corcommstyle{\bf\small\tt}

\def\corrl #1<<#2||#3>>{
\if\visiblecomments y
  \begin{quote} {\corcommstyle $<<$COMMENT$>>$ {\color{red}#1\marginpar{!!}}\\$<<$OLD$<<$} \end{quote}

{\color{red} 
 #2
 }

  \begin{quote} {\corcommstyle ==NEW== } \end{quote}
   \noindent\hrulefill
 
\vspace{-10pt} 
 
 \noindent\hrulefill
 
 \vspace{-10pt} 
 
 \noindent\dotfill
 
  #3
  
   \noindent\dotfill 

\vspace{-10pt} 
 
 \noindent\hrulefill
 
 \vspace{-10pt} 
 
 \noindent\hrulefill
  \begin{quote} {\corcommstyle $>>$END$>>$ } \end{quote}
 \else
  #3
 \fi
}

\long\def\longcorrl #1<<#2||#3>>{
\if\visiblecomments y
  \begin{quote} {\corcommstyle $<<$COMMENT$>>$ {\color{red}#1\marginpar{!!}}\\$<<$OLD$<<$} \end{quote}
 
 {\color{red}

  #2
  
  }
  
  \begin{quote} {\corcommstyle ==NEW== } \end{quote}
  
    \noindent\hrulefill
 
\vspace{-10pt} 
 
 \noindent\hrulefill
 
 \vspace{-10pt} 
 
 \noindent\dotfill
 
  #3
  
   \noindent\dotfill 

\vspace{-10pt} 
 
 \noindent\hrulefill
 
 \vspace{-10pt} 
 
 \noindent\hrulefill
  \begin{quote} {\corcommstyle $>>$END$>>$ } \end{quote}
 \else
  #3
 \fi
}

\def\mlabel #1
{
  \if\visiblecomments y
     \marginpar[\flushright \bf \footnotesize #1]{\bf \footnotesize #1}
  \fi
}

\def\flabel #1
{
  \if\visiblecomments y
       \hbox{\bf\footnotesize #1}
  \fi
}

\def\corrq #1<<#2>>{
\if\visiblecomments y
  \begin{quote} {\corcommstyle $<<$COMMENT$>>$ {\color{red}#1}\marginpar{!!}\\$<<$BEG$<<$} \end{quote}
  \noindent\hrulefill
 
\vspace{-10pt} 
 
 \noindent\hrulefill
 
 \vspace{-10pt} 
 
 \noindent\dotfill

  #2
 
  \noindent\dotfill 

\vspace{-10pt} 
 
 \noindent\hrulefill
 
 \vspace{-10pt} 
 
 \noindent\hrulefill 
  \begin{quote} {\corcommstyle $>>$END$>>$ } \end{quote}
 \else
  #2
 \fi
}

\long\def\longcorrq #1<<#2>>{
\if\visiblecomments y
  \begin{quote} {\corcommstyle $<<$COMMENT$>>$ #1\marginpar{!!}\\$<<$BEG$<<$} \end{quote}
  \noindent\hrulefill
 
\vspace{-10pt} 
 
 \noindent\hrulefill
 
 \vspace{-10pt} 
 
 \noindent\dotfill

  #2

  \noindent\dotfill 

\vspace{-10pt} 
 
 \noindent\hrulefill
 
 \vspace{-10pt} 
 
 \noindent\hrulefill 
  \begin{quote} {\corcommstyle $>>$END$>>$ } \end{quote}
 \else
  #2
 \fi
}

\def\corrc #1<<>>{
\if\visiblecomments y
  \begin{quote} {\corcommstyle $<<$COMMENT$>>$ \color{red} #1\marginpar{!!}} \end{quote}
\fi
}

\def\corre #1<<#2||#3>>{
\if\visiblecomments y
  #3\marginpar{\corcommstyle #1}
 \else
  #3
 \fi
}

\long\def\longcorre #1<<#2||#3>>{
\if\visiblecomments y
  #3\marginpar{\corcommstyle #1}
 \else
  #3
 \fi
}

\def\corrs #1<<#2||#3>>{
\if\visiblecomments y
  #3\marginpar{\corcommstyle #2 $\rightarrow$ #3\\ #1}
 \else
  #3
 \fi
}

\def\corro #1<<#2||#3>>{
#2}

\def\corrn #1<<#2||#3>>{
#3}

\long\def\longcorro #1<<#2||#3>>{
#2}

\long\def\longcorrn #1<<#2||#3>>{
#3}

\long\def\underconstruction #1<<<#2>>>{
\if\visiblecomments y
  \begin{quote} {\corcommstyle $<<$UNDER CONSTRUCTION - BEGIN$>>$ #1\marginpar{!!}} \end{quote}
  #2
  \begin{quote} {\corcommstyle $>>$UNDER CONSTRUCTION - END$>>$ } \end{quote}
 \else
 \fi
}

\def\showcomments{
  \let\visiblecomments y
}

\def\hidecomments{
  \let\visiblecomments n
}

\showcomments
\usepackage{amssymb}
\usepackage{soul}
\newcommand{\cn}{{\mathsf{CN}}}

\newcommand{\fn}{{\mathsf{FN}}}
\newcommand{\dfn}{{\mathsf{DFN}}}

\newcommand\hatR{\skew3\hat{R}}      

\newsavebox{\astrutbox}
\sbox{\astrutbox}{\rule[-5pt]{0pt}{20pt}}

\newcommand\shalf{\ensuremath{{\scriptstyle\frac{1}{2}}}}

\newcommand\nc{\newcommand}
\nc{\vect}[1]{\mbox{\boldmath $#1$}}

\begin{document}

\pagestyle{fancy}
\thispagestyle{plain}
\fancypagestyle{plain}{
\renewcommand{\headrulewidth}{0pt}
}

\makeFNbottom
\makeatletter
\renewcommand\LARGE{\@setfontsize\LARGE{15pt}{17}}
\renewcommand\Large{\@setfontsize\Large{12pt}{14}}
\renewcommand\large{\@setfontsize\large{10pt}{12}}
\renewcommand\footnotesize{\@setfontsize\footnotesize{7pt}{10}}
\makeatother

\renewcommand{\thefootnote}{\fnsymbol{footnote}}
\renewcommand\footnoterule{\vspace*{1pt}%
\color{cream}\hrule width 3.5in height 0.4pt \color{black}\vspace*{5pt}} 
\setcounter{secnumdepth}{5}

\makeatletter 
\renewcommand\@biblabel[1]{#1}            
\renewcommand\@makefntext[1]%
{\noindent\makebox[0pt][r]{\@thefnmark\,}#1}
\makeatother 
\renewcommand{\figurename}{\small{Fig.}~}
\sectionfont{\sffamily\Large}
\subsectionfont{\normalsize}
\subsubsectionfont{\bf}
\setstretch{1.125} 
\setlength{\skip\footins}{0.8cm}
\setlength{\footnotesep}{0.25cm}
\setlength{\jot}{10pt}
\titlespacing*{\section}{0pt}{4pt}{4pt}
\titlespacing*{\subsection}{0pt}{15pt}{1pt}

\fancyfoot{}
\fancyfoot[LO,RE]{\vspace{-7.1pt}\includegraphics[height=9pt]{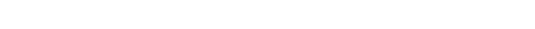}}
\fancyfoot[CO]{\vspace{-7.1pt}\hspace{13.2cm}\includegraphics{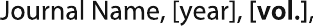}}
\fancyfoot[CE]{\vspace{-7.2pt}\hspace{-14.2cm}\includegraphics{head_foot/RF}}
\fancyfoot[RO]{\footnotesize{\sffamily{1--\pageref{LastPage} ~\textbar  \hspace{2pt}\thepage}}}
\fancyfoot[LE]{\footnotesize{\sffamily{\thepage~\textbar\hspace{3.45cm} 1--\pageref{LastPage}}}}
\fancyhead{}
\renewcommand{\headrulewidth}{0pt} 
\renewcommand{\footrulewidth}{0pt}
\setlength{\arrayrulewidth}{1pt}
\setlength{\columnsep}{6.5mm}
\setlength\bibsep{1pt}

\makeatletter 
\newlength{\figrulesep} 
\setlength{\figrulesep}{0.5\textfloatsep} 

\newcommand{\topfigrule}{\vspace*{-1pt}%
\noindent{\color{cream}\rule[-\figrulesep]{\columnwidth}{1.5pt}} }

\newcommand{\botfigrule}{\vspace*{-2pt}%
\noindent{\color{cream}\rule[\figrulesep]{\columnwidth}{1.5pt}} }

\newcommand{\dblfigrule}{\vspace*{-1pt}%
\noindent{\color{cream}\rule[-\figrulesep]{\textwidth}{1.5pt}} }

\makeatother

\twocolumn[
  \begin{@twocolumnfalse}
{\includegraphics[height=30pt]{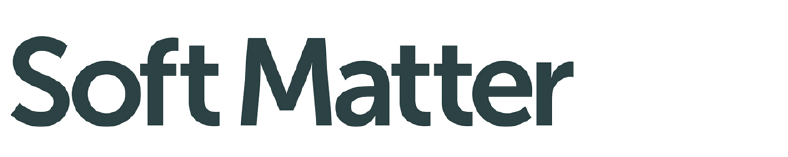}\hfill\raisebox{0pt}[0pt][0pt]{\includegraphics[height=55pt]{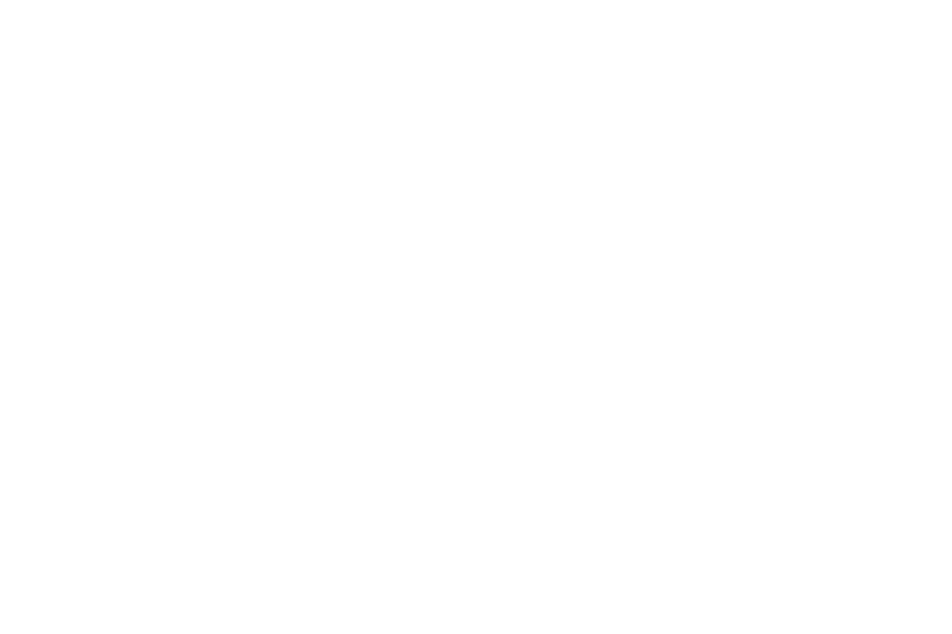}}\\[1ex]
\includegraphics[width=18.5cm]{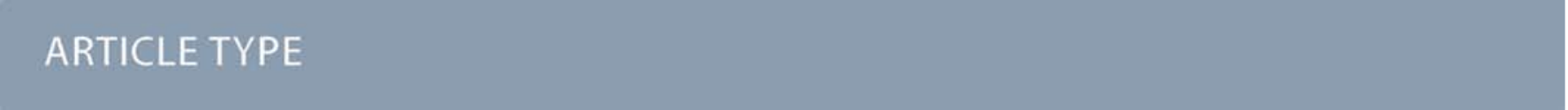}}\par
\vspace{1em}
\sffamily
\begin{tabular}{m{4.5cm} p{13.5cm} }

\includegraphics{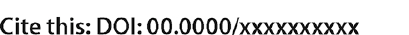} & \noindent\LARGE{\textbf{On intermittency in sheared granular systems }} \\
\vspace{0.3cm} & \vspace{0.3cm} \\

 & \noindent\large{Miroslav Kram\'ar,\textit{$^{a}$} Chao Cheng,\textit{$^{b}$} Rituparna Basak,\textit{$^{b}$} and Lou Kondic$^{\ast}$\textit{$^{b}$}} \\

\includegraphics{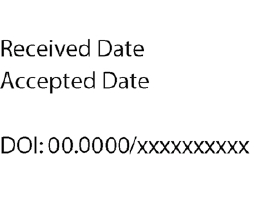} & \noindent\normalsize{Abstract 
We consider a system of granular particles, modeled by two dimensional frictional soft elastic disks, that is exposed to externally applied time-dependent 
shear stress in a planar Couette geometry. We concentrate on the external forcing that produces intermittent dynamics of stick-slip type. 
In this regime, the top wall remains almost at rest until the applied stress becomes sufficiently large, and then it slips.  We focus on the evolution of the system as it approaches a slip event. 
Our main finding is that there are two distinct groups of measures describing system behavior before a slip event.  The first group consists of global measures defined as system-wide averages at a fixed time. Typical examples of  measures in this group are  averages  of the normal or tangent forces acting between the particles, system size and number of contacts between the particles.  These measures do not seem to be sensitive 
to an approaching slip event. On average, they tend to increase linearly with the force pulling the spring.  The second group consists of the time-dependent measures that quantify the evolution of the system on a micro (particle) or mesoscale. Measures in this group first quantify the temporal differences between two states and only then aggregate them to a single number. For example, Wasserstein distance quantitatively  measures the changes of the force network as it evolves in time while the number of broken contacts quantifies the evolution of the contact network.  The behavior of the  measures in the second group changes dramatically before a slip event starts. They increase  rapidly  as a slip event approaches, indicating a significant increase in fluctuations of the system before 
a slip event is triggered. 

} \\

\end{tabular}
 \end{@twocolumnfalse} \vspace{0.6cm}     ]


\renewcommand*\rmdefault{bch}\normalfont\upshape
\rmfamily
\section*{}
\vspace{-1cm}


\footnotetext{\textit{$^{a}$~Department of Mathematics, University of Oklahoma, 601 Elm Avenue, Norman, OK 73019, USA. E-mail: miro@ou.edu }}
\footnotetext{\textit{$^{b}$~Department of Mathematical Sciences, New Jersey Institute of Technology, Newark, New Jersey 07102, USA. E-mail: kondic@njit.edu}}





\section{Introduction}

 Avalanches are phenomena that are well known on geological scales, with many familiar examples from earthquakes to snow avalanches and landslides.  The distribution of the times between the avalanches and their sizes is statistically similar to the distribution of abrupt events observed in many other systems where the consequences are less spectacular but crucial to understanding material responses that are of significant technological importance.   Relevant systems include dry and wet granular systems, suspensions, colloids, foams, yield-stress fluids, glass-forming materials, and several other soft matter systems relevant to our everyday life. The response of these systems to external driving is the subject of active research, with a large body of research considering an intermittent response where a system evolves via the stick-slip type of dynamics,  see~\cite{arcangelis_physrep_2016} for a review. Until recently, research concerning the predictability of upcoming events (slip, or avalanche) has been of a statistical nature.~\cite{daub_carlson_2010,kawamura_2012}. In recent years, new approaches based on
 the information emerging from simulations~\cite{staron2002preavalanche,welker2011precursors} or
 experiments~\cite{nerone2003instabilities,aguirre2006rearrangements,scheller2006precursors,zaitsev_epl_2008,daniels_hayman_2008,gibiat2009acoustic,johnson_geophys_2013,crassous_pre_2013}, in some cases coupled with machine learning approaches~\cite{tordesillas_mech_2018,tordesillas_remote_2019,liu_prx_2021}, have been considered. Despite progress, our ability to predict upcoming slips is still limited. Thus, it is important to devise more precise predictions, or at least to find out what type of information about the considered system is needed to make such predictions feasible.

The intermittent type of dynamics of granular systems is often explored because it provides a good testing ground for various theoretical approaches, see~\cite{luding2021jamming} for a recent review. In particular, both experiments and simulations provide detailed  information about particles as well as their interactions. Experimentally, the interactions between the particles have been extensively analyzed by using methods based on photoelasticity~\cite{daniels_hayman_2008, Hayman2011_pag,clark_prl12,tordesillas_bob_pre12,zadeh2019crackling,zadeh2019enlightening,kozlowski_pre_2019}. 
Complementary information can be obtained for a wide variety of systems~\cite{peters05,ciamarra_prl10,arcangelis_iop11,griffa_2011,ferdowsi_pre_2014,kumar2014macroscopic,
carlevaro_pre_2020,luding_softmatter_2022} by simulations based on discrete element methods. Therefore, a significant amount of research has been carried out and here we mention only a few examples. Various statistical measures of intermittent dynamics  have been considered in detail in simulations~\cite{ciamarra_prl10,arcangelis_iop11,walker_pre12,griffa_2011,ferdowsi_pre_2014}, and experiments with particular focus on quantifying intermittency were reported as well~\cite{Kaproth2013_sci,harth2020intermittent}. The connections between different systems experiencing intermittency have been discussed extensively~\cite{dahmen_prl09,dahmen_group15,dahmen_natcom_2016}  and significant progress has been reached in understanding how a system yields, in particular based on the shear transformation zone concept~\cite{daub_carlson_2010,maloney_pre_2006,falk_langer_2011}. 

One important question when considering intermittent dynamics is the relation between 
micro (particle scale) and macro (system size) behavior. The interactions between the particles can be captured by a force network. Visual inspection of these networks  shows that local interactions between the particles give rise to organized structures that form spontaneously on  a mesoscale (with a typical length scale of ten or so particle diameters).  By now, it is widely accepted that the force networks play an important role in determining the system-wide response.  However, it is still not clear how to extract relevant information about the intermittent dynamics from the properties of these networks, or some other particle-scale/mesoscale properties of the system.

Our main goal is to identify measures that exhibit a clear change in their trend as a system approaches a slip event, since finding such measures is the first and necessary step towards developing the ability to forecast slip events. 
To achieve this goal, we consider a simple system in two dimensional (2D) planar Couette geometry, with the top boundary
pulled by a harmonic spring, see Fig.~\ref{fig:system_figure}. We investigate a wide variety of measures that quantify 
various properties of inter-particle contacts, forces (force networks), and particles' dynamics. Some of the measures are 
classical while others, based on persistent homology (PH),  have been implemented recently~\cite{epl12,kondic_2016,pre13,pre14,physicaD14,Pugnaloni_2016}. 
We will show that global measures, obtained by averaging any of the considered quantities over the whole system, do not change their behavior as the 
system approaches a slip event. On the contrary, the measures quantifying the micro and mesoscopic evolution of the system on a  short time scale start 
increasing nonlinearly well before the onset of the slip. Thus, these measures show significant potential for predicting an upcoming slip event.  
 Our interpretation of this finding is that local time-dependent measures capture increasing fluctuations on micro and mesoscopic scales that lead to a slip event. The reported results are  for 2D systems. This choice reduces  the computational cost as well as difficulties associated with a large amount of data. This being said, we note that the considered methods and measures easily extend to 3D.

The rest of this manuscript is structured as follows. Section~\ref{sec:simulations} focuses on 
the description of simulation techniques.  In Sec.~\ref{sec:def} we
define the measures that we consider in this paper.  Section~\ref{sec:results} 
provides the main results, motivated by consideration of a single 
slip event that is discussed in Sec.~\ref{sec:single}.  In Sec.~\ref{sec:stat} we start by
providing an overview of the statistical analysis that we carry out, applied
to the global measures in Sec.~\ref{sec:global}, and 
then to the local ones in Sec.~\ref{sec:local}. Section~\ref{sec:conclusions} is
devoted to the summary and discussion of future directions.  Animations of DEM simulations
as well as of some of the considered measures are presented in Supplementary Materials.

\section{Simulations}
\label{sec:simulations}

\begin{figure}[t!]
    \centering
    \includegraphics[width=0.9\linewidth]{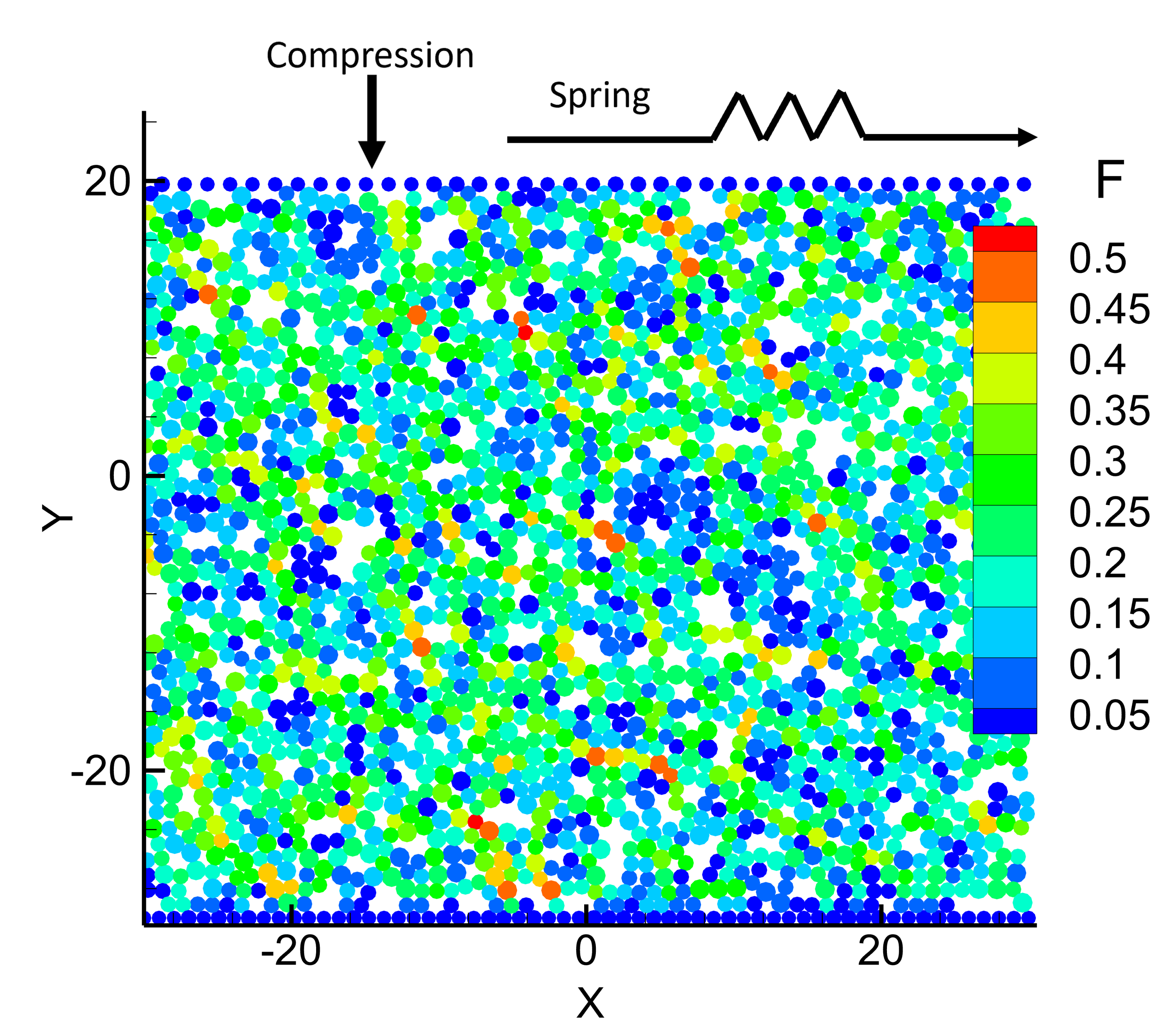}
    \caption{Snapshot of the considered system.  The system particles are colored by their total current  force magnitude while the wall particles are shown in blue.   The top wall is exposed to a linear force, in the $+x$ direction, by a spring and (constant) pressure force is applied in the $-y$ direction.  The bottom wall is fixed. }
    \label{fig:system_figure}
\end{figure}

In this section, we provide a short overview of the considered simulations. We also explain how we store the relevant information about the system for the subsequent analysis. Since the simulation techniques are identical to the methods described in Ref.~\cite{kovalcinova_scaling} we limit ourselves to a summary and refer the reader to
earlier work~\cite{dijksman_2018} for the description of experiments carried out with photoelastic particles that provided the needed material parameters.  In our simulations, we 
model granular particles as 2D soft frictional disks, and place $N = 2500$ disks (system particles)
between two horizontal rough walls (made up of wall particles) placed parallel to the horizontal $x$ axis, 
see Fig.~\ref{fig:system_figure}. 
The system particles are bi-disperse, with 25\% of large particles and 75\% of small particles, and the diameter of a large particle is 25\% larger than that of a small particle.
The top wall is made of small particles spaced slightly apart from each other. This spacing and increased friction of the wall particles (as discussed in what follows), reduces substantially the slip of the system particles along the wall. The bottom wall is also made of small particles. Since there is no observable slip of the system particles next to this wall, we leave the wall particles at a distance equal to the particle diameter. 
The bottom wall is kept fixed,  while the top one is pulled by a harmonic spring moving with the velocity $v_s$ in the $+x$ direction.  The left-right boundary conditions are periodic; we have carried out limited simulations with wider (in the $x$ direction) domain to confirm that the width of the domain does not influence the results.  The influence of the domain height (in the $y$ direction) is briefly mentioned in Sec.~\ref{sec:results}.  

We use the linear spring-dashpot model to 
describe the interactions between the system particles and between system and wall particles.  In what follows, 
we use the diameter of small particles, $d$, as the length scale, their mass, $m$, as the mass scale,
and the binary collision time, $\tau_c$, as the time scale.  
Motivated
by experiments with photoelastic particles~\cite{dijksman_2018}, we use 
$d=1.27$ cm, $m=1.32$ g, and $\tau_c = 1.25\times10^{-3}$ s, as appropriate for  particles of 
Young modulus of $Y\approx 0.7$ MPa. 
For such value of $\tau_c$, the normal 
spring constant is $k_n = m\pi^2/2{\tau_c}^2 \approx 4.17$ N/m,   
and the tangential spring constant (needed for modeling of 
tangential forces using the Cundall-Strack model~\cite{cundall79}) is $k_t = 6k_n/7$, which is close to the value 
used previously \cite{goldhirsch_nature05}.  The coefficient of static friction, $\mu$, is equal to $0.7$ for 
particle-particle contacts and $\mu = 2$ for particle-wall contacts. We choose the  larger value for 
the latter to further reduce slipping of particles adjacent to the walls, as discussed above. 
The force constant of the spring applied to the top wall, $k_s$, is significantly smaller than the one 
describing particle interactions, $k_s = k_n/400$. The (constant) restitution coefficient is 0.5.
In addition, a normal compression force is applied in the 
$-y$ direction to  model an externally applied pressure (force/length in 2D) of $p=0.02$; gravitational effects 
are not included.  We note that with our choice of units, the numerical value of the applied pressure is of the 
same order of magnitude as the average overlap (compression) of the particles.  

It is well known that a sufficiently large $p$ and sufficiently small spring 
speed, $v_s$, are needed for the system to  enter a stick-slip regime~\cite{ciamarra_prl10}. We found by 
experimenting that for $p = 0.02$ the value $v_s =  1.5\times10^{-3}$ is appropriate to 
induce stick-slip dynamics.  We integrate Newton’s equations of motion for both the translational and 
rotational degrees of freedom using a fourth-order predictor-corrector method with time step $dt = 0.02$. 
The states of the system, used to compute the quantities presented in this paper, are stored every 10 time steps, so 
$\tau_c/5$ apart.  All the results are presented using $\tau_c$ as the time scale. 

\begin{figure}[t!]
\centering
\includegraphics[width=0.8\linewidth]{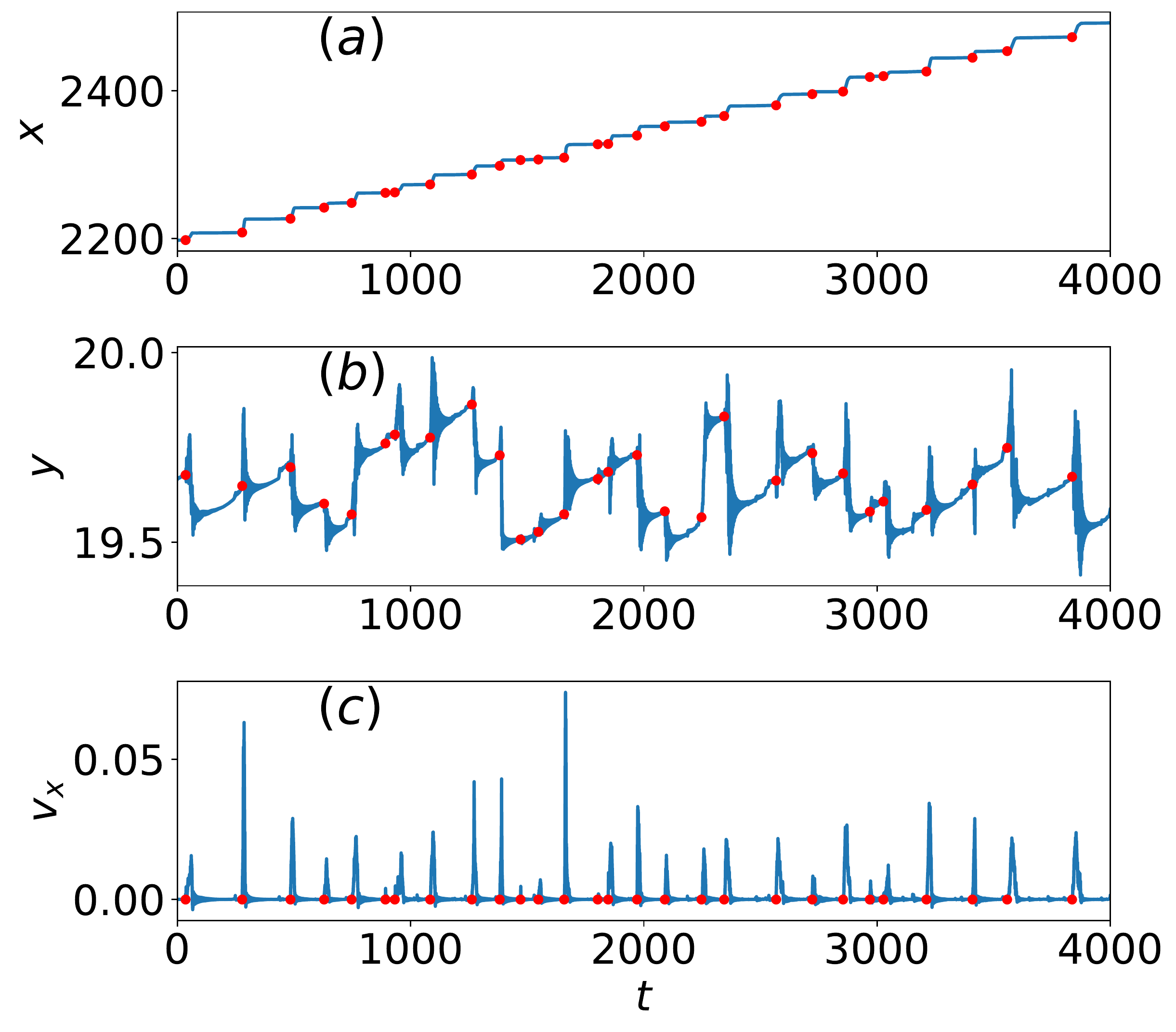}
\caption{Representative system evolution from the beginning of the production runs, as described in the text.  The time (horizontal) axis in this and the following figures is given in units of $\tau_c$. (a) $x$ position of the top wall, (b) $y$ position of the top wall, and (c) top wall velocity, $\vx$, in the shearing direction. The red dots indicate the times at which our algorithm detects the beginning of the slip events.  }
\label{fig:stick-slip}
\end{figure}

The simulation protocol starts by applying a pressure $p$ to the top wall and then letting the system 
relax until the ratio of kinetic/potential energy becomes sufficiently 
small.  To ensure that the particles have settled we require that this ratio drops below $10^{-5}$. The results are not sensitive to small changes of this value.  After the particles are settled, we start 
moving the spring in the $+x$ direction. Initially, the wall remains almost stationary. Due to bulk compression of 
granular particles, the wall  moves slightly  in the direction of applied spring force, as discussed later in the text. Once the spring force 
becomes sufficiently large, the top wall starts to slip.  Initially, the  particles' rearrangements cause 
the $y$-position of the wall to  decrease gradually. To avoid this transient regime, 
we shear the system for approximately $6\times 10^5$ time steps to ensure that a steady state is reached. It is the long-time average of the $y$ coordinate of the top wall becomes constant.  After this preparation stage, we start production runs. We discard additional  $3 \times 10^4$ time steps to ensure once again that the system is in steady state, and then start collecting data.
Figure~\ref{fig:stick-slip}(a-c) shows a short time window of the wall positions in the $x$ and $y$ direction as well as the speed of the wall in the shearing direction, $x$.
Figure~\ref{fig:stick-slip_zoomedin} depicts the detailed behaviour during the first slip event, shown in Fig.~\ref{fig:stick-slip}, to illustrate typical dynamics.  We note in passing that each slip is accompanied by a jump of the wall 
in the $y$ direction, see Fig.~\ref{fig:stick-slip}(b)
and Fig.~\ref{fig:stick-slip_zoomedin}(b).  
To reach reasonable statistics, we carry out simulations for a long time so that a large number (400-500) 
of slip events occur.  The total number of data points (extracted every $\tau_c/5$) is $3\times 10^5$. We note that good temporal resolution of a slip event shown in 
Fig.~\ref{fig:stick-slip_zoomedin} suggests that this sampling rate is is sufficiently large to allow 
for precise detection of slip events. 

\section{Definitions of relevant quantities}
\label{sec:def}

In this section, we present the measures that we use to analyze the behavior of the system before the onset of a slip event. We start by defining what we mean by the start and end of a slip event in Sec.~\ref{sec::slip_detection}. In Sec.~\ref{sec::FN} we formalize the notion of the force networks and differential force networks that we use to study the evolution of the system. To quantify the structure of these networks and their evolution we use persistent homology (PH), which is a valuable tool of topological data 
analysis, see~\cite{mischaikow, edelsbrunner-harer} for a review, 
and~\cite{physicaD14,hiraoka_pnas2016} for examples of applications.  We will briefly explain the important concepts behind PH in Sec.~\ref{sec::PH}. Finally, in Sec.~\ref{sec::contacts} we summarize the quantities that we use to characterize contacts between the particles.

\subsection{Slip detection}
\label{sec::slip_detection}

\begin{figure}[t!]
\centering
\includegraphics[width=0.8\linewidth]{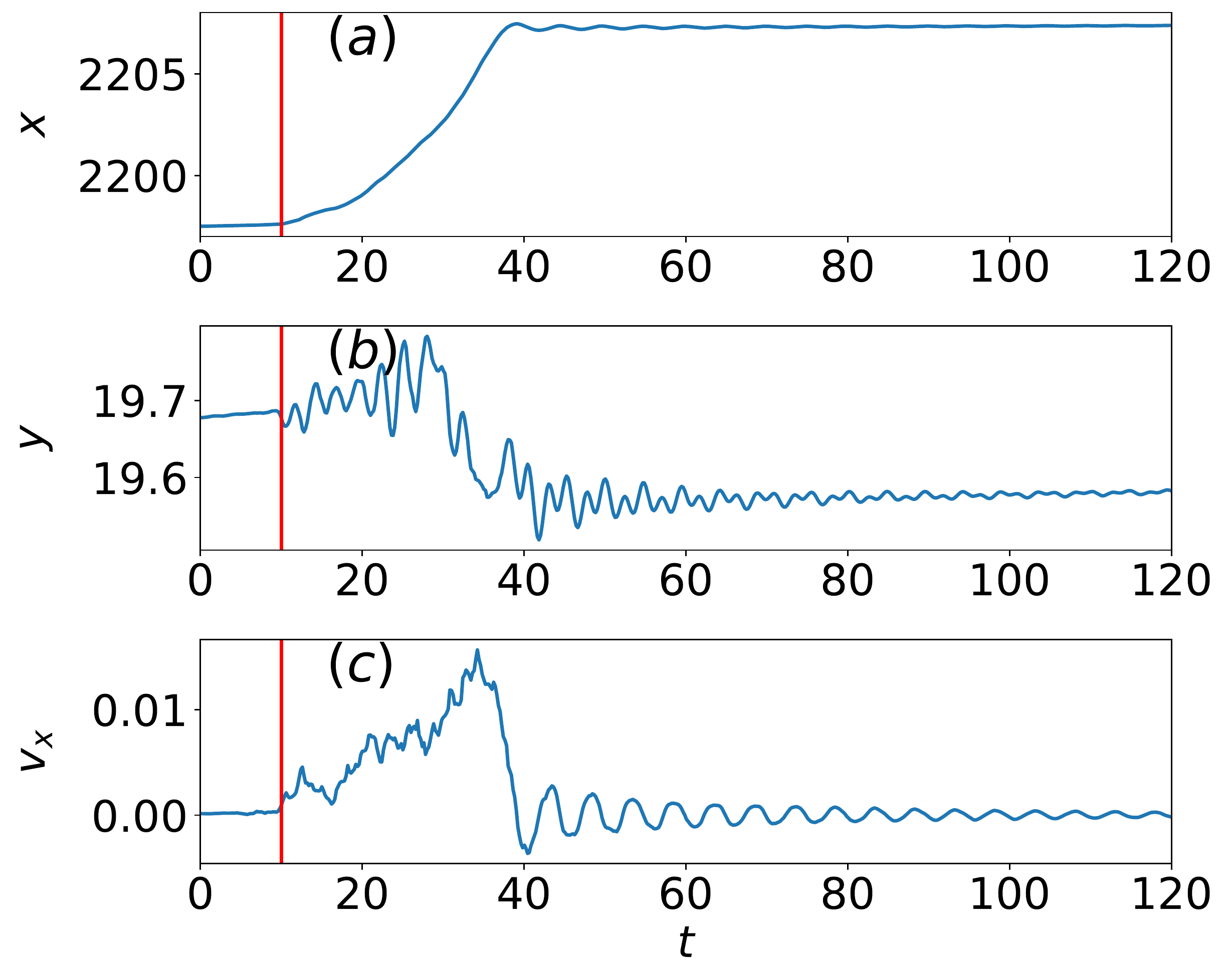}
\caption{Wall behavior around the first slip event depicted in Fig.~\ref{fig:stick-slip}. The red line indicates the time $t_0$, defined in Sec.~\ref{sec::slip_detection}, at which the slip starts and the behavior of considered measures changes abruptly.
}
\label{fig:stick-slip_zoomedin}
\end{figure}

To study the behavior of the system before the onset of a slip,  we need to properly determine when the wall starts sliding and the system enters 
the slip regime. To make sure that we precisely identify the times at which the wall starts moving, while ignoring small oscillations of the wall velocity 
after a slip event (see Fig.~\ref{fig:stick-slip_zoomedin} (b-c)), we use the following protocol that consists of two parts.   
These two parts could be labeled as `rough' and `fine'.  The `rough' 
part is based on the fact that the top wall velocity, $\vx$, tends to differ by several orders of magnitude in the slip and stick 
phases. We choose a threshold $v_f^*  = 2\times 10^{-3}$ which is an order of magnitude smaller than a typical velocity during a slip event and at least an order of magnitude larger than the average velocity during the stick phase (which is nonzero in part due to bulk compression of granular particles, caused by the spring force).  
The rough part of the algorithm identifies the times at which $\vx$ crosses the value $v_f^*$ as the beginnings of individual slip events.  To avoid considering small oscillations following a slip event as separate events, we require that all slip events are separated by more than $30\tau_c$.  In the second part, we fine-tune the starting times of the slip events as follows.  First, we compute the mean, $\bar{v}_x$, and standard deviation,  $s_{\vx}$, of $\vx$ for all times at which $\vx < v^*_f$. These values are used to define a new threshold $v_f = \bar{v}_x + s_{\vx}  \approx 2.5\times10^{-4}$. Finally, we adjust the times identified as the beginning of slip events, in the `rough' part of the algorithm,  by decreasing them until the value of $\vx$ drops below  $v_f$.  We have verified that the precise values of the thresholds have only a minor influence on the results that follow
Naturally, a much smaller value of the `rough' threshold would lead to detection of large number of `microslip' events (see, e.g.~\cite{long_granmat_2019}) characterized by small fluctuations in the wall position. In the present work, we focus on large slip events only.

For the results that follow, the identification of the end of a slip is not necessary. However, we provide a protocol that is designed to avoid premature detection of a slip end due to small oscillations following a slip. To achieve this we require that the wall velocity  $\vx$ at the end of the slip event, as well as its average over the preceding  $50$  states, is smaller than $v_f^*$. The chosen number of states roughly corresponds to the period of small oscillations following a slip event.

\subsection{ Force networks and differential force networks}
\label{sec::FN}

The available information about particle interactions can be encoded by a force network. In this paper, we utilize a force network defined by normal interparticle forces,  however, force networks defined by tangential (or total) forces could be considered as well. 

We start by defining a time dependent contact network, $\cn(t)$, that describes which particles are in 
contact at a given time $t$. This network is based on vertices $V_c(t) = \{v_i\}_{i=1}^N$ that correspond to 
the centers of system particles, $\{p_i\}_{i=1}^N$ (we do not consider wall particles).  An edge $\langle v_i, v_j\rangle$  belongs to $\cn(t)$ if the particles $p_i$ and $p_j$ are in contact at time $t$.

\begin{figure}[t!]
\centering
\includegraphics[width=0.48\linewidth]{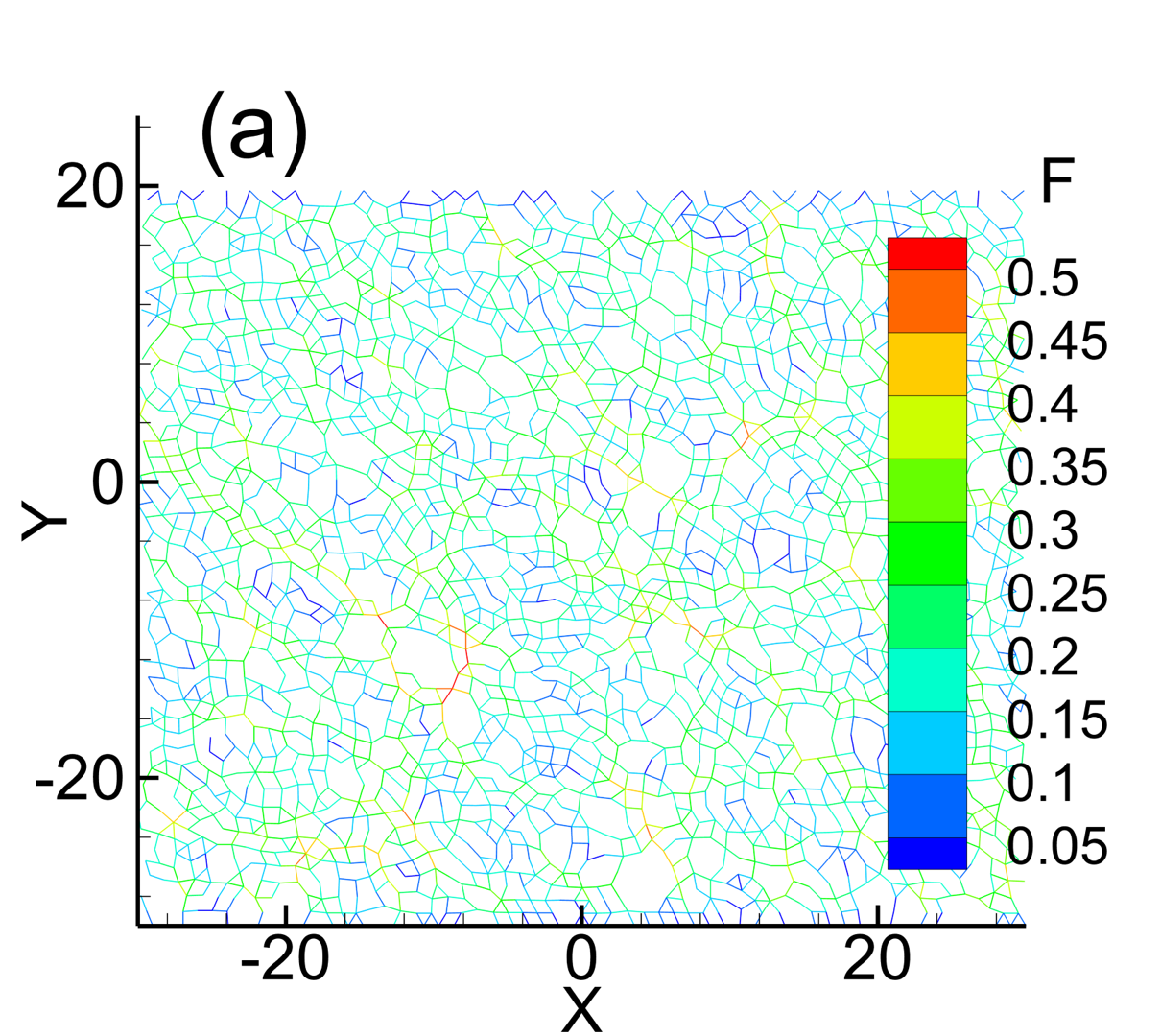}
\includegraphics[width=0.48\linewidth]{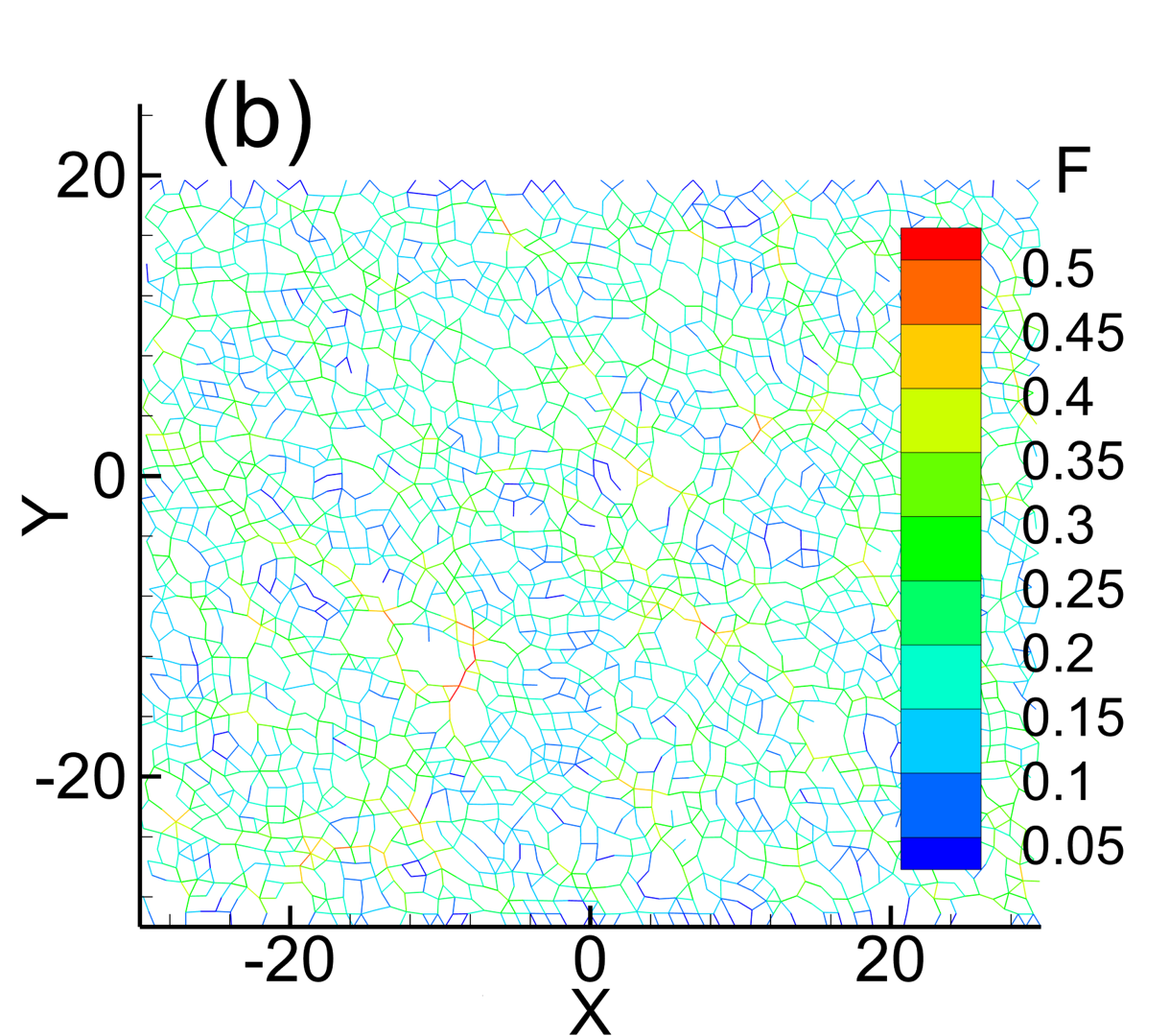}
\includegraphics[width=0.48\linewidth]{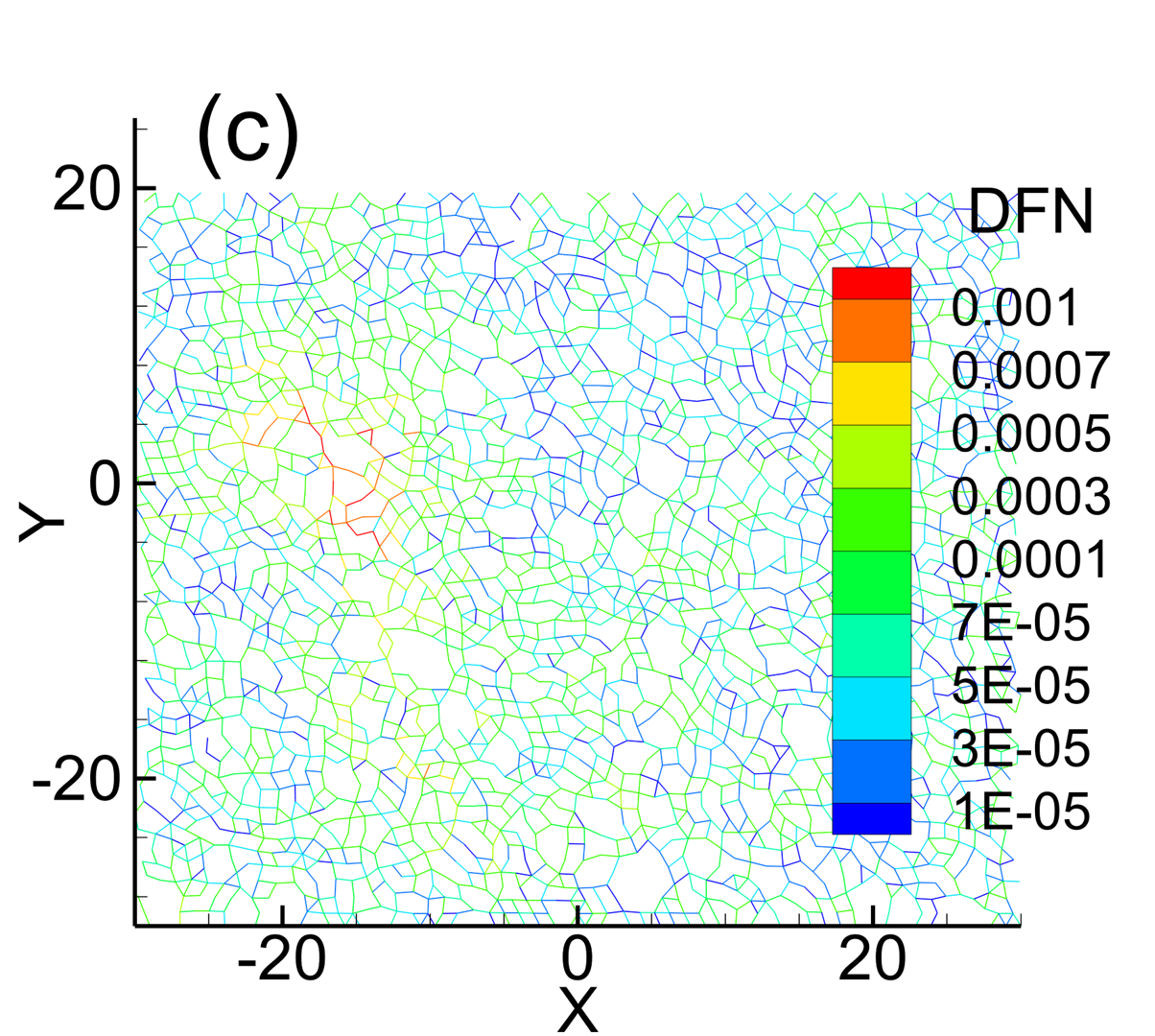}
\includegraphics[width=0.48\linewidth]{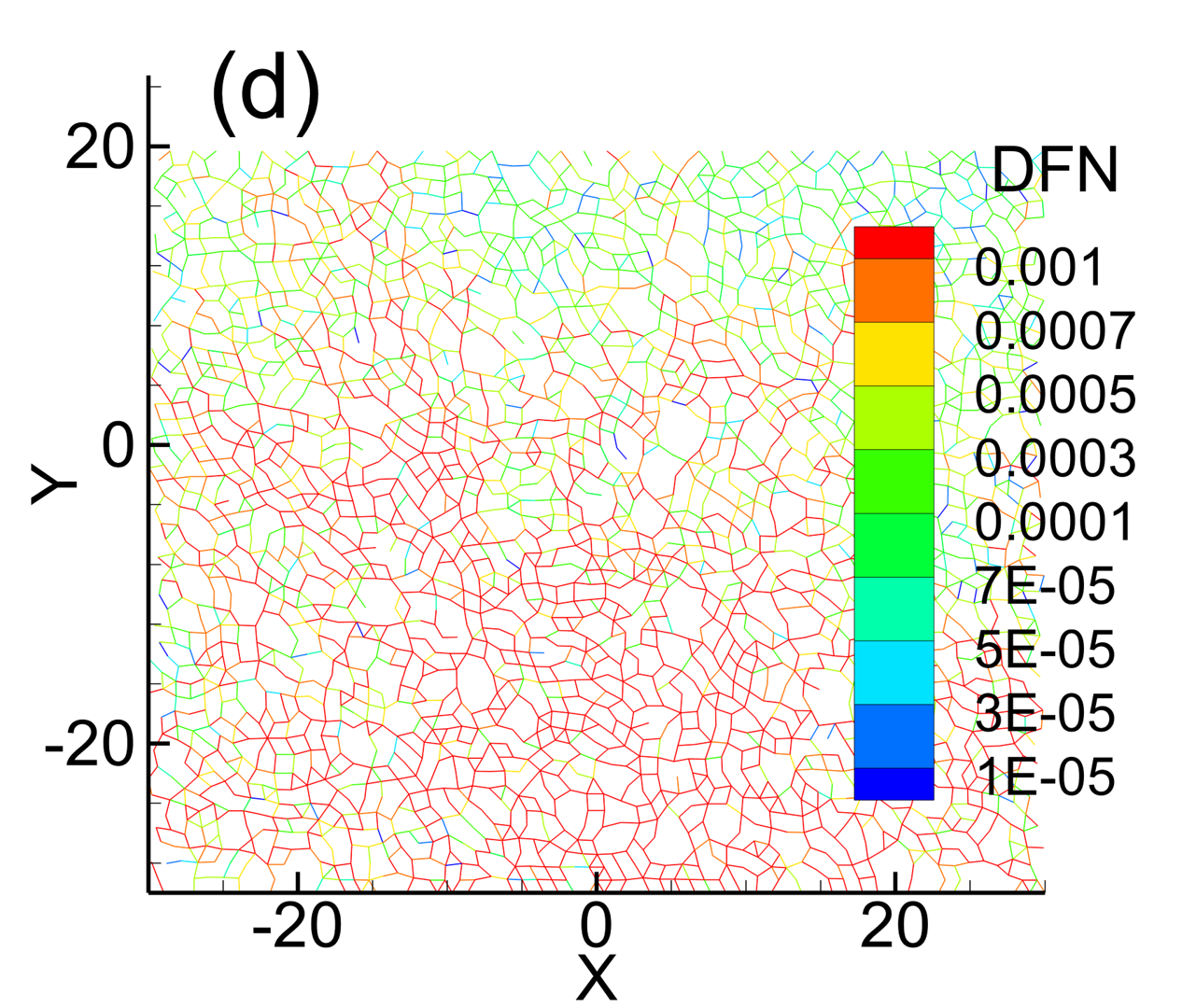}
\caption{Force and differential force networks for the slip event, shown in Fig.~\ref{fig:stick-slip_zoomedin}.
(a) Force network before the onset of a slip event at time $t = t_0 - 4$ where $t_0$ is the time at which the slip event starts.
(b) Force network at the onset of a slip event at $t = t_0$, (c) Differential force network at  $t = t_0 - 4$, and 
(d) Differential force network at $t = t_0$.
Note that the changes of the network (shown in (c-d)) 
are on the scale that is much smaller than  a typical force in the force network shown 
in (a-b). Thus,  the networks in (a-b) appear almost identical. Animations of the differential force networks are available as Supplemental Materials.}
\label{fig:force_network}
\end{figure}

The force network, $\fn(t)$, is defined by assigning weights to the edges of $\cn(t)$, so that the weight of the edge $\langle v_i, v_j\rangle$ is the magnitude of the normal force acting between the particles $p_i$ and $p_j$.  Figures~\ref{fig:force_network}(a,~b) show the force  network at two different times before a slip event,
depicted in Fig.~\ref{fig:stick-slip_zoomedin}, occurs. The time evolution of the force network is illustrated by the animations in Supplementary Materials. 

One possibility for encoding differences between the force networks $\fn(t)$ and $\fn(t +\tau_c/5)$ is 
to consider a differential force network, $\dfn(t)$, that expresses how much the force network changes between two considered time instances. Because the force bearing contacts might be created or destroyed during the time interval $[t, t+\tau_c/5]$, the edges of $\dfn(t)$ are given by the union of the edges in $\cn(t)$ and $\cn(t+\tau_c/5)$. The weights of the edges in  $\dfn(t)$, defined as the absolute value  of the difference of its weights in $\fn(t)$ and $\fn(t+\tau_c/5)$, 
indicate the changes in forces acting between the particles  between  time $t$ and $t+\tau_c/5$.  If an  edge is not present in $\fn(t)$ or $\fn(t+\tau_c/5)$, then its weight in the corresponding force network is set to zero. Figures~\ref{fig:force_network}(c) and (d) show two differential force networks. The weights of the edges in $\dfn(t)$ increase 
as the system approaches a slip event, see also Supplementary materials for animations. 
Figure~\ref{fig:force_network}(c) shows a localized increase of the weights of the $\dfn$  at $\approx (-4,0)$. A further research is necessary to investigate the relation between such localized changes and localized particles modes discussed in~\cite{maloney_prl_2004} and the references therein.
Here we just point out that the localized changes in $\dfn$ do not necessarily involve particle dynamics.

To study $\dfn(t)$ we compute the maximum force value, $f^*$, such that the edges of $\dfn(t)$ with weights larger than $f^*$  percolate through the system.  In 
what follows, we will consider both left-right and top-bottom percolation,  and call the  corresponding $f^*$ values $f_{plr}$ and $f_{ptb}$, respectively. 

\subsection{Measures derived using persistent homology}
\label{sec::PH}

To identify complex structures exhibited by force networks we use persistent homology (PH) which is one of the major tools of topological data analysis. PH has been extensively used to describe complex patterns  in a variety of 
settings~\cite{hiraoka_pnas2016, hirata_science2013, kramar_physD_2016, taylor2015topological}. There is also a growing body of literature that uses PH to study granular systems exposed to compression~\cite{epl12,pre14}, vibrations~\cite{ardanza_pre14,kondic_2016,Pugnaloni_2016},
or shear~\cite{gameiro_prf_2020}.  In this section, we only provide a brief summary of PH and introduce the measures used in this paper. Detailed guidance for using PH to analyze force networks is available elsewhere~\cite{physicaD14}.

Every weighted network in two spatial dimensions can be represented by two persistence diagrams, $\pd$s, that provide a compact but informative description of the structure of this network. Each diagram is a collection of points in a plane and these points describe how the topology of sub-graphs containing only the edges with weights exceeding some threshold $T$ changes as the value of $T$ is decreased. The first persistence diagram, $\pd_0$, encodes the structure of connected components that can be related to so-called `force chains'. To be more precise, every point $(b,d) \in \pd_0$ corresponds to a connected component that appears at the threshold $T = b$ and merges with another connected component for $T = d$. Hence, the lifespan of the point $(b,d)$, given by $b-d$, indicates the prominence of the connected component corresponding to this point. Similarly, a point $(b,d) \in \pd_1$ indicates that a loop appears in the sub-graph for $T = b$. Once a loop appears at $T = b$ it is present in all the sub-graph for all $T<b$. In this paper we follow the convention introduced in~\cite{physicaD14} and set $d=0$ for every point in $\pd_1$. Supplementary materials include an animation of the force network and the corresponding $\pd$s for a selected slip event. 

Information contained in $\pd$s can be further compressed in several ways. One possibility of compressing a $\pd$ to a single number is to compute the sum of the lifespans of all the points in the diagram. We call this 
quantity total persistence, $\tp0$ or $\tp1$ depending on whether it is extracted from 
$\pd_0$ or $\pd_1$.  In previous studies $\tp$ was found to be a very useful 
quantity e.g., in~\cite{gameiro_prf_2020} $\tp1$ was correlated with the viscosity of a 
shared suspension, showing directly the connection between force network properties and 
rheological properties of the considered system.

In addition to considering system-wide averages at individual times (such as $\tp$), we also use PH to quantify the {\it time evolution} of the force network. The space of persistence diagrams is a complete metric space for a variety of metrics~\cite{edelsbrunner-harer}. The main idea behind defining a metric on this space is to match the points in one diagram with the points in the other. This matching can be done in different ways leading to different metrics. In this paper 
we consider the matching that minimizes the sum (W2 distance)
of the squares of $L_\infty$  distances between the matched points. If the $W2$ distance is computed between the $\pd_0$ ($\pd_1$) diagrams 
corresponding to the force networks $\fn(t)$ and $\fn(t+\tau_c/5)$, we denote it by W2B0 (W2B1). For simplicity we suppress the time argument on W2's.  
The distance depends on the difference between the times
at which the compared  $\pd$'s are computed.  We have verified that during the stick phases this dependence is 
approximately linear, confirming that the considered sampling rate is large enough to resolve the temporal evolution of the force network~\cite{kramar_chaos_2021}.

\subsection{Broken, mobile and nonmobile contacts}
\label{sec::contacts}
In this section, we define the measures that we use to study the changes of the contact network $\cn(t)$ and the nature of the contacts. 
If the contact between the particles $p_i$ and $p_j$ is present at time $t$, but 
not at $t + \tau_c/5$, we say that the contact is broken.  A simple measure to quantify the  difference between $\cn(t)$ and  $\cn(t+\tau_c/5 )$ 
is given by the ratio,  $r_{bc}(t)$, between the number of broken 
contacts and the total number of contacts at time $t$. If the contact between two particles disappears, then the force 
previously acting between them vanishes and we refer to this force as a broken force.  The average broken force, $f_{bc}$, is defined as the 
sum of all broken forces divided by the number of broken contacts.  

Until now, we only discussed quantities based on the normal component of the force acting between the particles.  
In  Sec.~\ref{sec:results} we also discuss two quantities that involve both 
the magnitude, $F_t$, of the tangential component of the force and the magnitude, $F_n,$ of the normal component.  One considered
quantity is simply the ratio, $F_{t}/F_{n}$,
calculated separately for each contact and then averaged over all contacts.  
 The other related quantity is the ratio of mobile to non-mobile contacts, RMN. A contact between two particles is mobile if the 
 ratio $F_{t}/F_{n}$ is at the Coulomb threshold, $\mu$. Note that within the implemented model $F_t \leq \mu F_n$, so that mobile 
 contacts are the ones for which ${F_t}/{F_n}$ reaches the largest possible value. In our computations, a contact is 
 considered mobile if ${F_t}/{F_n} > \mu - \epsilon$, with $\epsilon = 10^{-3}$. All other contacts are called non-mobile.  As
 a reminder, we do not consider particle-wall contacts.

\section{Results}
\label{sec:results}

\subsection{Motivation: A single slip event} 
\label{sec:single}
To motivate the following discussion, we first discuss a 
single slip event that starts at time $t_0$. We are interested in the behavior before the start of this event.  Thus, we present the results in terms of  $\Delta t = t -t_0$.

\begin{figure}[t!]
    \centering
    \includegraphics[width = 0.8\linewidth]{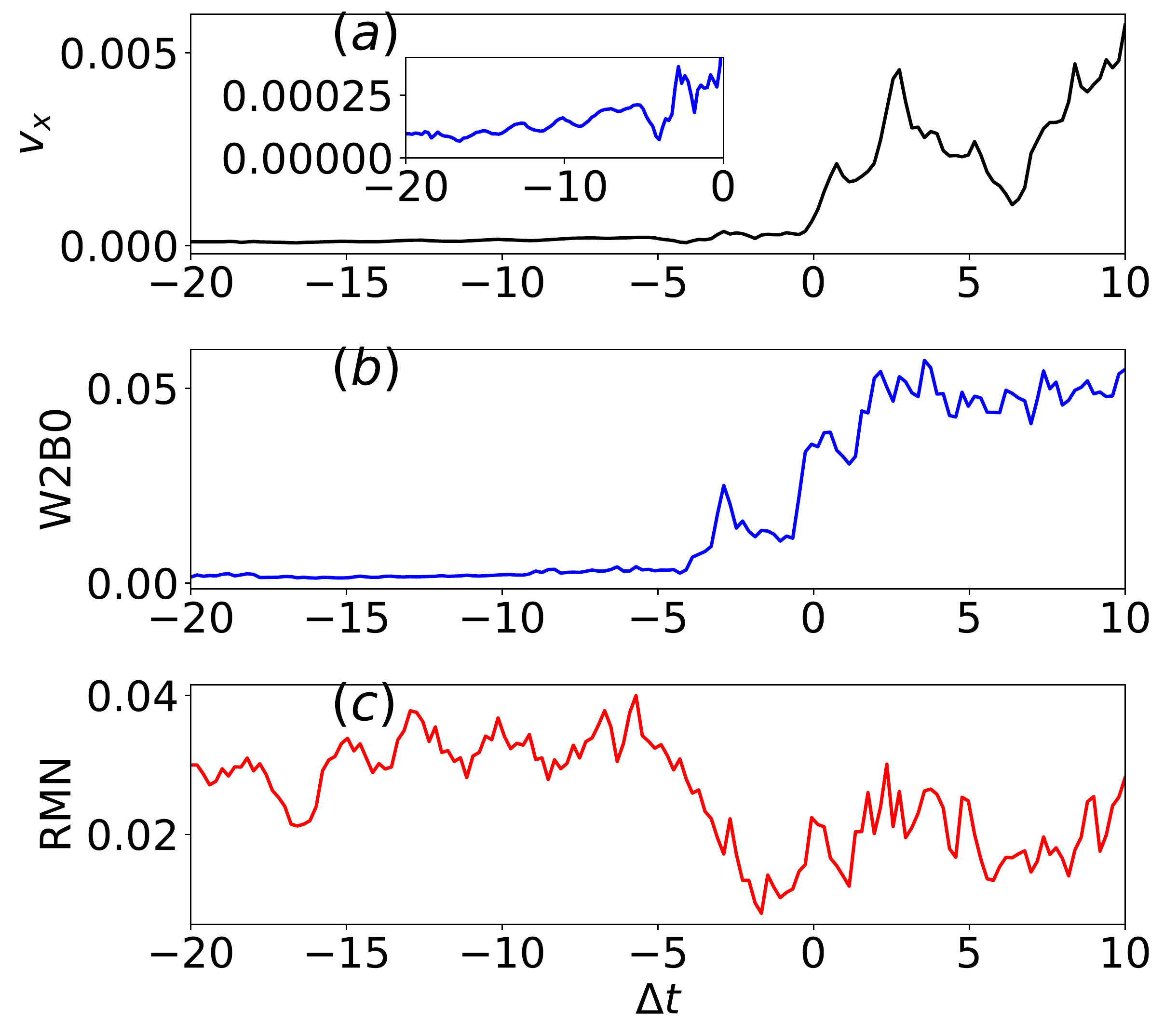}
    \caption{(a) Velocity of the top wall, $\mbox{v}_x$,  in the $x$ direction for the slip event shown  in Fig~\ref{fig:stick-slip_zoomedin} and Fig.~\ref{fig:force_network}. The inset shows the behaviour before the beginning of the slip. The small increase of $\vx$ before the slip is typical but other features such as the local minimum  around $-4$ are not generic.  The value $\Delta t = 0$ indicates the  time at which the event starts.  (b) W2B0, (c) ratio of mobile to non-mobile contacts, RMN. }
    \label{fig:event_curves}
\end{figure}

Figure \ref{fig:event_curves}(a-c) shows the velocity of the top wall, $\vx$, W2B0,  and the ratio of the mobile and non-mobile contacts, RMN. 
Note that due to our definition of $t_0$  the wall is essentially at rest for $\Delta t < 0$ (except the slow drift due to bulk compression combined with small fluctuations).  
However, Fig.~\ref{fig:event_curves}(b) shows that 
W2B0, which measures changes in the structure of the force network, increases significantly 
already around $\Delta t = -4$.  This indicates that the force network starts changing
rapidly as the system approaches a  slip event.  The fact that the W2B0  detects this increased 
activity suggests its potential for predicting slip events.  On the contrary, RMN does not 
exhibit any clear trend. This finding motivates a more careful statistical analysis of a large number of slip events. 
Such analysis, presented in what follows, demonstrates that the difference between 
W2B0 and RMN, shown in Fig.~\ref{fig:event_curves}, is not just a coincidence.

\subsection{Statistical Analysis}
\label{sec:stat}

\begin{figure}
    \centering
    \includegraphics[width = 0.8\linewidth]{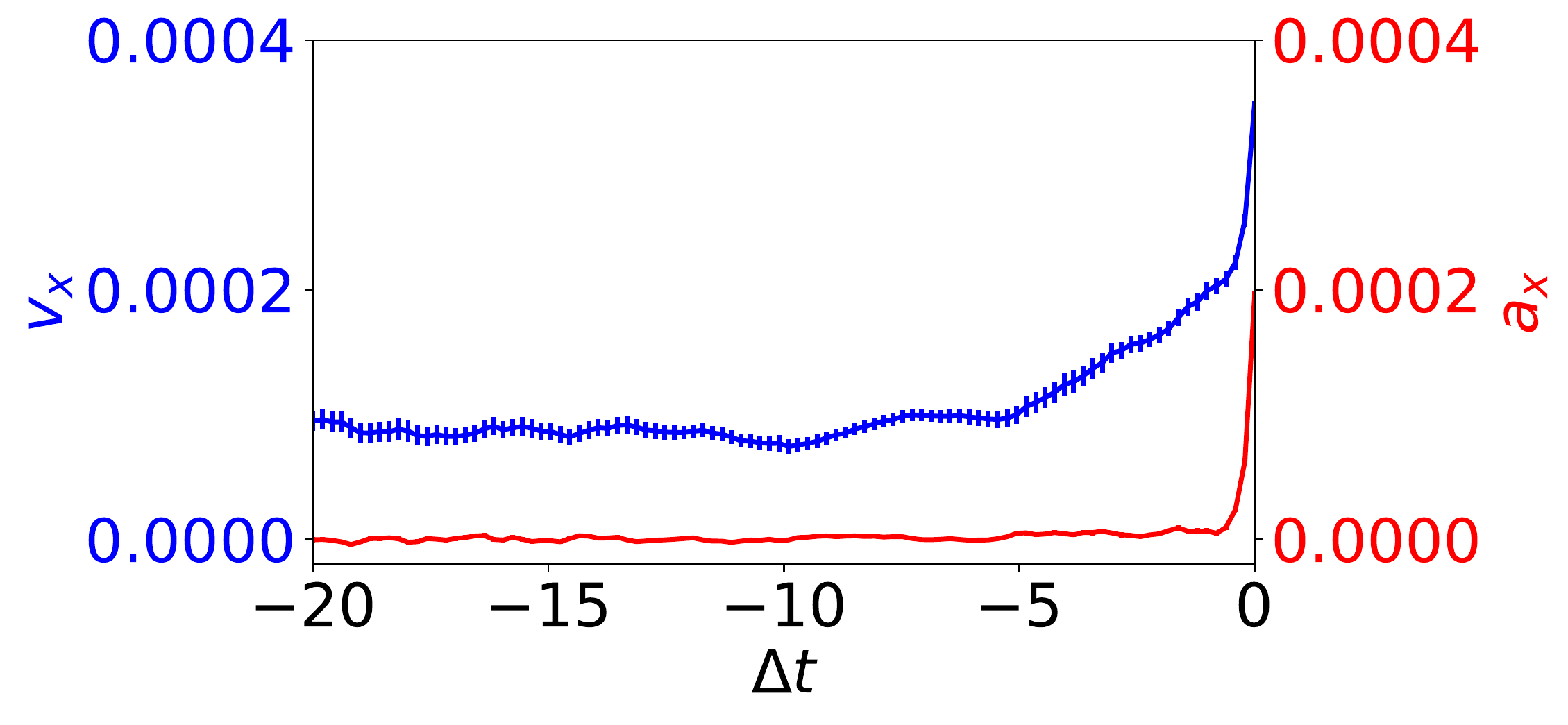}
    \caption{The mean of the wall speed, $\vx$,  and acceleration, $a_x$ before slip events, averaged over all slip events. 
    The value $\Delta t=0$ corresponds to the beginning of the slip events, $t_0$. In this and the following figures, the means are 
    calculated using the approach described at the beginning of Sec.~\ref{sec:stat}, and the error 
    bars show the standard errors.} 
    \label{fig:dv_v}
\end{figure}

To study the behavior of the system over a large number of slip events,  we average the considered measures over different slips as follows. For each event we record the individual measures for the last $100$ consecutive samples before its beginning, that is for $\Delta t \in\{ -20, -19.8, \ldots, 0\}$. Then, we compute  average of each measure for fixed values of $\Delta t$.  

Figure~\ref{fig:dv_v} shows the mean of wall velocity, $\vx$, and of its acceleration, averaged over the 
complete set of $465$ slip events. Due to our definition of a slip, the wall os almost stationary  before the slip starts at $\Delta t = 0$ (small  component of the wall velocity before the slip is 
due to bulk compression as mentioned earlier in the text)~\footnote{For further physical insight, note that for the present choice of parameters, 
dimensionless speed of $1.0\times 10^{-4}$ correspond to the physical value of $\approx 1.0\times 10^{-5}$ m/s.}. However, there is a change in the trend around $\Delta t = -5$. For $\Delta t < -5$ the mean $\vx$ fluctuates while it increases steadily for  $\Delta t > -5$, however this increase is orders of magnitude smaller than the increase observed at the beginning of the slip. 
Note that the mean roughly doubles in the time interval $[-5, -0.2]$, and then it almost doubles again between
two consecutive outputs as the slip starts.  

We proceed by discussing two sets of different measures. The first set consists of the measures that are obtained as  (global) system-wide averages of the considered 
properties of the system at a given time. The second set of measures is devised to assess micro and mesoscale changes that occur as the system evolves in time.  
We will show that these two sets of measures provide very different information about the system's behavior before a slip event.

\begin{figure}[t!]
    \centering
    \includegraphics[width=\linewidth]{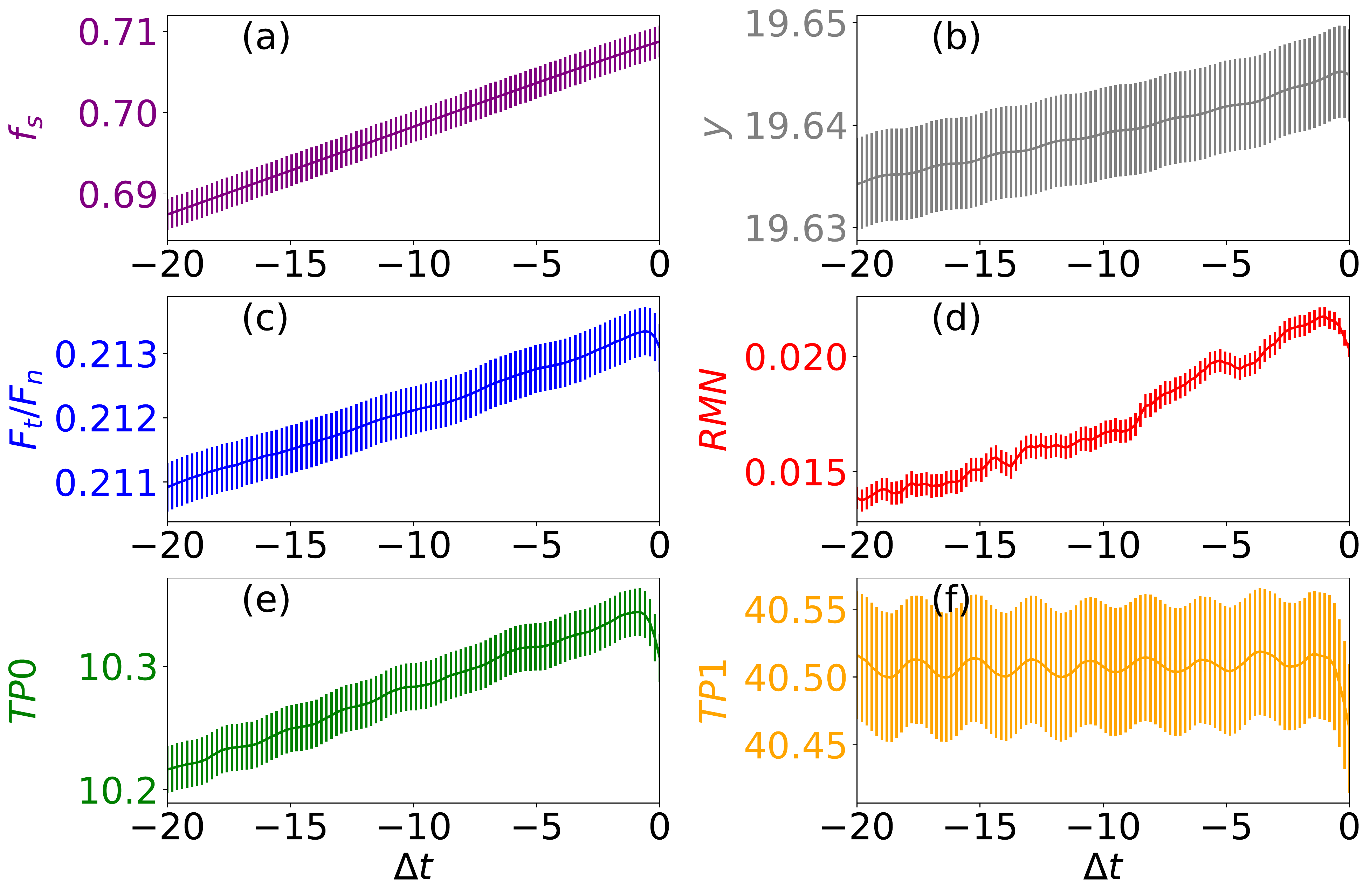}
    \caption{Global measures averaged over all slip events: 
        (a) spring force, $f_s$, (b) $y$ position of the top wall, (c) the ratio $F_t/F_n$, (d) the ratio of mobile to non-mobile contacts, RMN, (e) TP0, (f) TP1. }
    \label{fig:static_pre_slip}
\end{figure}

\subsubsection{Global measures}
\label{sec:global}

Figure~\ref{fig:static_pre_slip} shows our results for the set of measures based on the system-wide
averages.  Averaging the considered measures over a large number of events produces reasonably smooth results, despite the large variability of these quantities between individual slip events. 
For brevity, we do not discuss this variability in more quantitative terms here. In the rest of this paper, we only report the standard errors to indicate how well the individual means are estimated. 

Figure~\ref{fig:static_pre_slip}(a) depicts the force pulling the top wall. This measure is not 
strictly speaking a system-wide average but we show it  since the results are similar to other 
quantities presented in this figure.  The value that this force reaches, before individual slips, varies considerably (see, e.g.~\cite{ciamarra_prl10,Denisov2017_sr} for examples of simulations in similar settings).  This variability shows the stochastic nature of stick-slip type dynamics. Nevertheless, Fig.~\ref{fig:static_pre_slip}(a) indicates that, 
despite large variations, the mean of the force pulling the top wall is well estimated and increases linearly.

Figure~\ref{fig:static_pre_slip}(b)  shows that the mean of the $y$ coordinate of the wall position also increases 
with time, since the system expands as the force applied by the  spring increases.  This effect is known as 
Reynolds dilatancy (could be also interpreted in terms of the Poisson ratio of the granular system considered)~\footnote{The 
connection between these two interpretations formulates an interesting question which we however do not discuss here in order
to keep the discussion focused.}
and is caused by 
the systems response to an increased applied stress. Note that the effect is very weak and the system only expands by a small fraction of the particle diameter. Based on the present data, it is difficult to confirm quadratic increase of the wall position with applied shear stress discussed recently~\cite{ren_prl_2004}.  We note that this (weak) expansion of the system also leads to a small decrease of another global measure, the contact number, shown in Appendix, Fig.~\ref{fig:contactnumber}.

Figures~\ref{fig:static_pre_slip}(c,~d) show two related quantities defined in Sec.~\ref{sec:def}. The ratio of tangential and normal forces, $F_t/F_n$,  and the ratio of mobile to nonmobile contacts (RMN).  The value of $F_t/F_n$ increases linearly with the applied spring force. This is not surprising because, in a static system, one expects $F_t$ to counteract the applied force and thus the linear dependence seems natural. While this argument is more appropriate in a limit of a single layer of separate particles between the walls, Fig.~\ref{fig:static_pre_slip}(c) shows that it holds, at least on average, for the system considered here as well. 
The increase of RMN is a consequence of the increase of tangential forces, ignoring for the moment the fact that the normal force may change as well. It is worth pointing out that these simple scaling results hold only on average and the behavior of the individual quantities can be very different for a single event, see Fig.~\ref{fig:event_curves}. We also note that the total normal force between
the particles increases as slips are approached, making the above argument only approximate.  

The next two global, system-wide measures are derived from persistence diagrams. 
Figure~\ref{fig:static_pre_slip}(e) shows the total persistence for connected components, TP0, which increases
linearly.  This is not surprising since this measure is expected to scale with the applied force. The behavior of TP1, 
shown in Fig.~\ref{fig:static_pre_slip}(f) is different and deserves further attention.  
Recall that TP1 is the sum of the lifespans of the points in the $\pd_1$. The birth coordinate of a point in $\pd_1$ is given by the magnitude of the weakest force in the loop corresponding to this point. Because the death coordinate is always zero, the lifespan of the point is equal to its birth coordinate. 
 The oscillations of the forces on the weakest links cause oscillations of TP1.  It turns out that oscillations of the forces on the weakest links are caused by minor oscillations of the top wall, barely visible in 
 Figure~\ref{fig:static_pre_slip}(b). These oscillations, in turn, are essentially damped aftershocks
 following slip events.  
Additional simulations (not shown here for brevity) of the systems for which the average distance between the top and bottom wall is approximately twice as large (carried out by doubling the 
number of particles) show that the period of these oscillations scales with the system height.  This finding suggests that the oscillations are caused by compression waves propagating through the system in the $y$ direction. 
We note that the components of the Cauchy stress tensor show similar behavior as the measures discussed so far. These components are shown in 
Fig.~\ref{fig:stresses} in the Appendix. 

\begin{figure}[t!]
    \centering
    \includegraphics[width=0.75\linewidth]{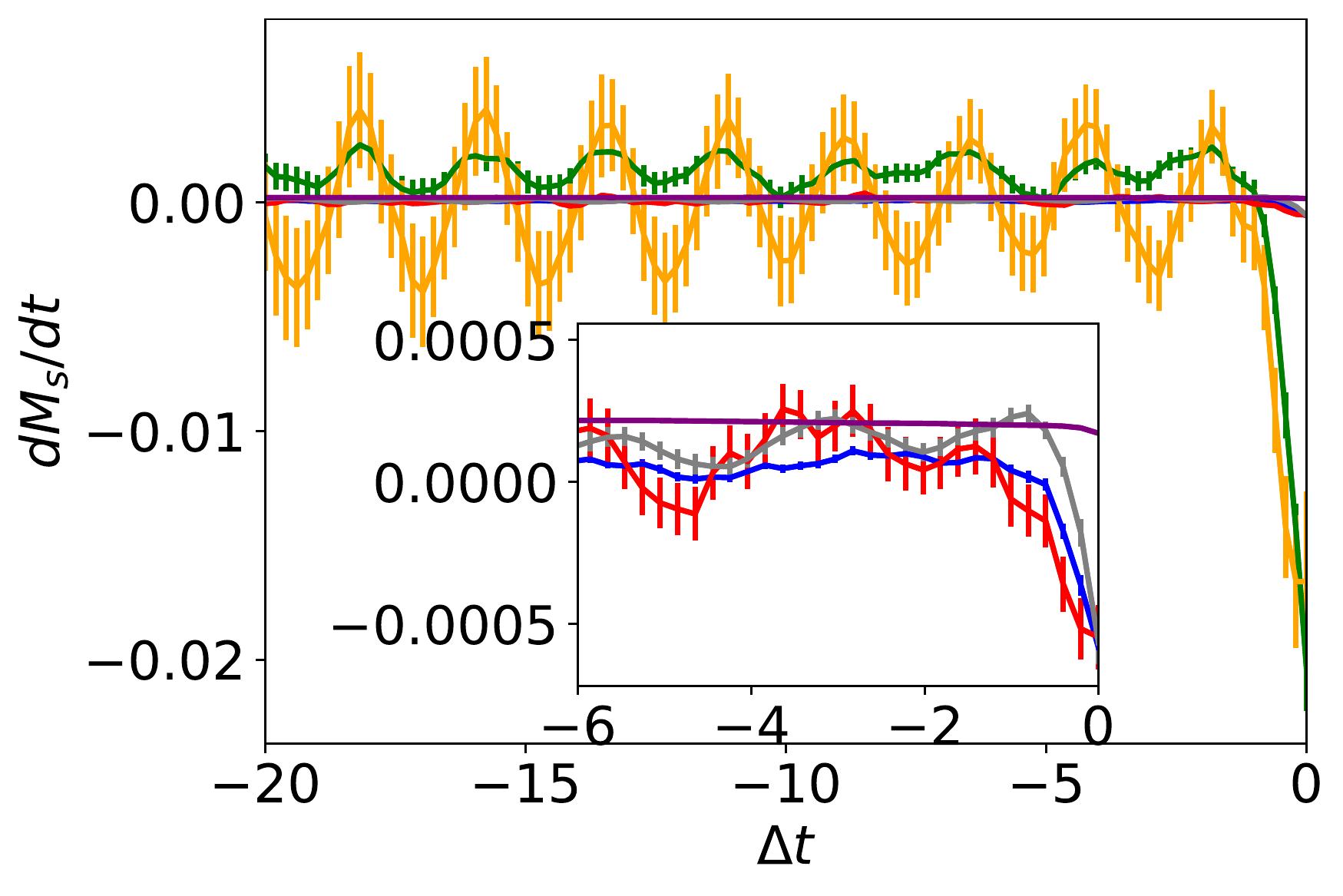}
    \caption{Derivatives of the global measures, shown in Fig.~\ref{fig:static_pre_slip},  are depicted using the  same colors.
    The inset zooms into the measures whose change is hardly visible in
    the main plot.
    }
    \label{fig:static_derivativs}
\end{figure}

To further show that the system-wide averages do not exhibit any clear signs of the approaching slip, as well as to facilitate comparison with the results presented in Sec.~\ref{sec:local}, we consider their derivatives. Figure~\ref{fig:static_derivativs}, which depicts the derivatives of the measures presented in Fig.~\ref{fig:static_pre_slip},  indicates that the derivatives do not change dramatically, except for a couple of data points right before a slip occurs. This behavior, very close to the beginning of a slip event is due to relaxation of the forces caused by the fact that the system starts slowly evolving and breaking up contacts between the particles. However, 
this relaxation only happens very close to the slip itself and is not particularly useful as a slip precursor.  
Hence, we conclude that global measures do not capture the behavior which could be useful for
predicting an imminent slip event. This motivates the need for measures that assess the evolution of the system on the micro (particle) and mesoscopic spatial scales. 

\subsubsection{Local measures}
\label{sec:local}

In  Sec.~\ref{sec:def}  we defined measures that are capable of quantifying the changes of the system on the particle scale as well as on mesoscopic scales relevant to our study of the evolution of the force network. We remind the reader that to quantify these changes we first compare local and mesoscopic differences between the consecutive samples and only then aggregate them to a single number quantifying the difference.  As discussed previously in the context of W2 distances, the output rate in our simulations
is sufficiently high so that the main features of the results are rate-independent.  

\begin{figure}
    \centering
    \includegraphics[width = \linewidth]{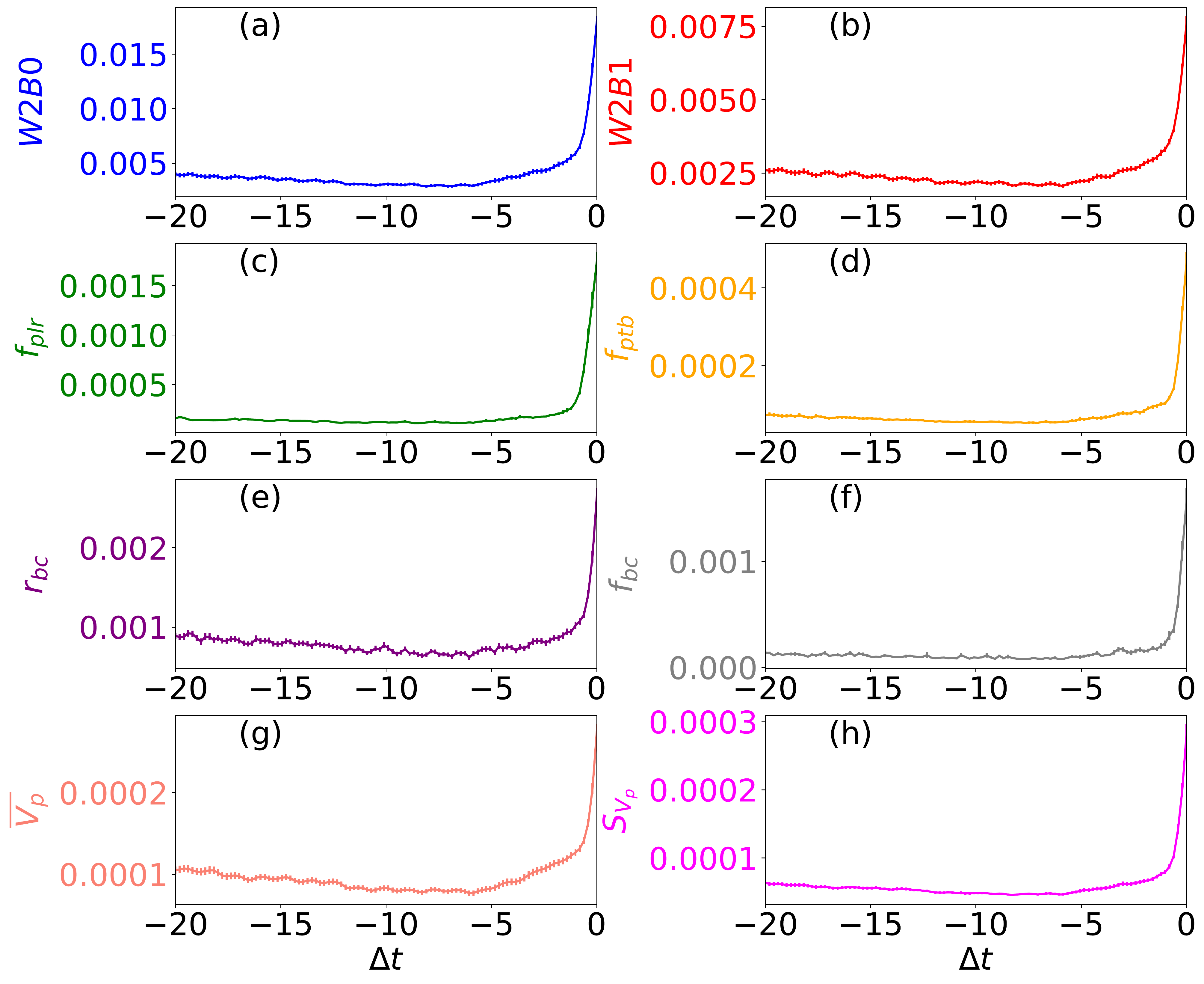}
    \caption{(a) W2B0 distance, (b) W2B1 distance, (c) the horizontal percolating differential normal force, $f_{plr}$, (d) the vertical percolating differential normal force, $f_{ptb}$, (e) the ratio of broken contacts to total contacts, $r_{bc}$, (f) the average normal force of broken contacts, $f_{bc}$, (g) the mean of system particle speed, $\overline{v_p}$, (h) the standard deviation of system particle speed, $s_{v_p}$. } 
    \label{fig:dynamic_pre_slip}
\end{figure}

Figure~\ref{fig:dynamic_pre_slip} shows the results, again averaged over all slip events.  
Figures~\ref{fig:dynamic_pre_slip}(a,~b) depict the W2B0 and W2B1 distances measuring the differences between two consecutive persistence diagrams $\pd_0$ ($\pd_1$) that capture the  structure of the connected components (loops) present in the force network.  The measures, shown in Fig.~\ref{fig:dynamic_pre_slip}(c,~d), also evaluate temporal changes in the force network structure by utilizing the notion of the differential force network. 

The parts (e,~f) focus on broken contacts:  part (e) shows the ratio of the number of broken contacts divided by the number of all contacts, and (f) shows the average normal force of these broken contacts. We note that the behavior  of the broken contacts  is very different from the time evolution of the number of contacts shown in Appendix, Fig~\ref{fig:contactnumber}. This stems from the fact that the number of contacts (static, global measure) is aggregated from a single state of the contact network  while the number of broken contacts is obtained by comparing two consecutive states.

Finally, the parts (g,~h) show the mean speed, $\bar{v}_p$, of the system particles, and its standard deviation, $s_{v_p}$. These quantities are calculated as $\bar{v}_p = \sum_{i=1}^Nv_{i}/N $ and 
$s_{v_p} = \sqrt{\sum_{i=1}^N(v_{i}-\bar{v}_p)^2/N} $, where $v_i$ is the speed of the $i$-th  particle.  Note that the ratio of kinetic to potential energy of the particles, see Fig.~\ref{fig:ek/ep} in the Appendix, shows similar generic behavior, although an increase of that measure appears to be delayed, compared to, eg., $\bar{v}_p$.

Although there is no obvious direct connection between the measures presented in Fig.~\ref{fig:dynamic_pre_slip}, they all show similar behavior. 
In particular, all of them start growing rapidly before the beginning of a slip event.  We recall that this behavior is similar to the behavior of the wall velocity, $\vx$, see Fig.~\ref{fig:dv_v}. Therefore,  
the trend exhibited by the dynamic local measures is completely different from the one exhibited by the static global measures, Fig.~\ref{fig:static_pre_slip}, or their derivatives, Fig.~\ref{fig:static_derivativs}. We note that the considered dynamic measures  exhibit a slow decrease before the onset of the slip.  We expect that this decrease is a consequence of a slow relaxation from the preceding slip event. However,  further research is necessary  to analyze this decrease in more detail. 

All measures shown in Fig.~\ref{fig:dynamic_pre_slip} quantify the time evolution of the system on either particle scale or mesoscale. The fact that these measures increase before the onset of a slip event indicates that the  evolution on both particle scale and mesoscale intensifies before the slip. However, this increased activity is not detected by the global measures. This suggests that the increased activity is limited to local fluctuations. We expect that  these fluctuations intensify until they eventually reach a critical level and trigger the slip event, which leads to global rearrangements.  Notice that these fluctuations do not only affect the  force networks but also the movement of the particles, as indicated by the increase of their velocities shown in  Fig.~\ref{fig:dynamic_pre_slip}(g~-~h).

Before closing this section, we comment  on the  very different behavior of the wall velocities in the $x$ and $y$ directions (compare Figs.~\ref{fig:dv_v} and~\ref{fig:static_derivativs}).  This difference is caused by the different nature of these two measures.  The velocity $v_y$ depends on the (global) pressure while the velocity $\vx$  depends on local interactions between the wall and the particles.

\section{Conclusions}
\label{sec:conclusions}

We consider a granular system exhibiting intermittent dynamics known as the stick-slip regime. 
To detect the slip events, we use a strict a posteriori method, based on the wall movement. Even though the system is 
essentially static before slip starts, we can identify measures that dramatically change their behavior as 
 a slip event is approached.

The measures that we consider in this paper fall into two categories.  The first category of global measures is obtained by averaging over the whole system.  Such measures include 
the system size (wall position), contact number, or system-wide measures of the normal and tangential forces between the particles.
We find that before a slip event these measures show (on average) approximately linear behavior. There is no clear change in their  behavior  almost until the slip starts. 
The second category of the measures quantifies local and mesoscopic changes of the system, 
computed based on information at different time instances.  
Such measures include the Wasserstein distance measuring time evolution of the
force network, percolating properties of the differential force network, 
or the number of broken contacts, among others. We find that the
measures in this category behave differently.  Namely,  they start to increase in a  nonlinear fashion well before a slip starts.  
The local nature of these measures suggests that their behavior is caused by spatial and temporal fluctuations in the system which cannot be detected by the global measures. 

We hypothesize that the intensity of the fluctuations in the system increases until it overcomes the stabilizing effects of the force network and triggers a slip event. Therefore,  information about the evolution of the system on micro and mesoscopic scales seems to be vital for accurately predicting the occurrence of a slip.

We expect that the local measures, introduced in this paper, could be used to predict an upcoming slip event. However, 
the variations between different slip events suggest that more advanced statistical methods involving either machine learning or some 
complementary approach will be needed to achieve this goal. The development of such methods will be the subject of our future work.

\section*{Appendix}

\begin{figure}
    \centering
    \includegraphics[width = .8\linewidth]{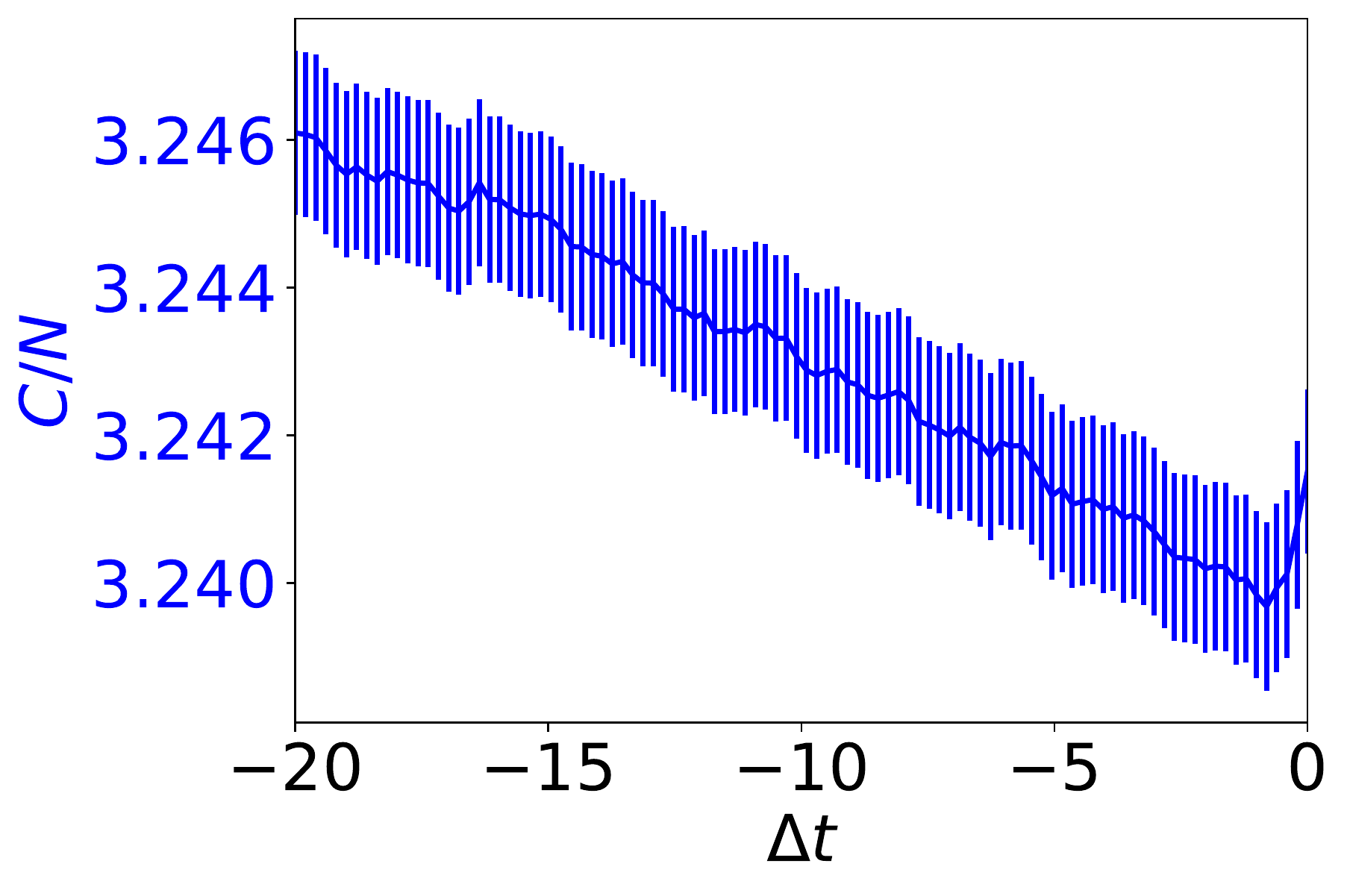}
    \caption{ Number of contacts per particle f averaged over all slip events.}
    \label{fig:contactnumber}
\end{figure}

We present some additional global and local measures that show similar trends as the measures presented in the main body of the paper. Figure~\ref{fig:contactnumber} shows the number of contacts $C$ per particle averaged over all the slip events. Note this this global static measure slowly decreases with time in an almost linear fashion. As the other global measures it only changes its behavior very close to the beginning of the slip. 

\begin{figure}
    \centering
    \includegraphics[width = \linewidth]{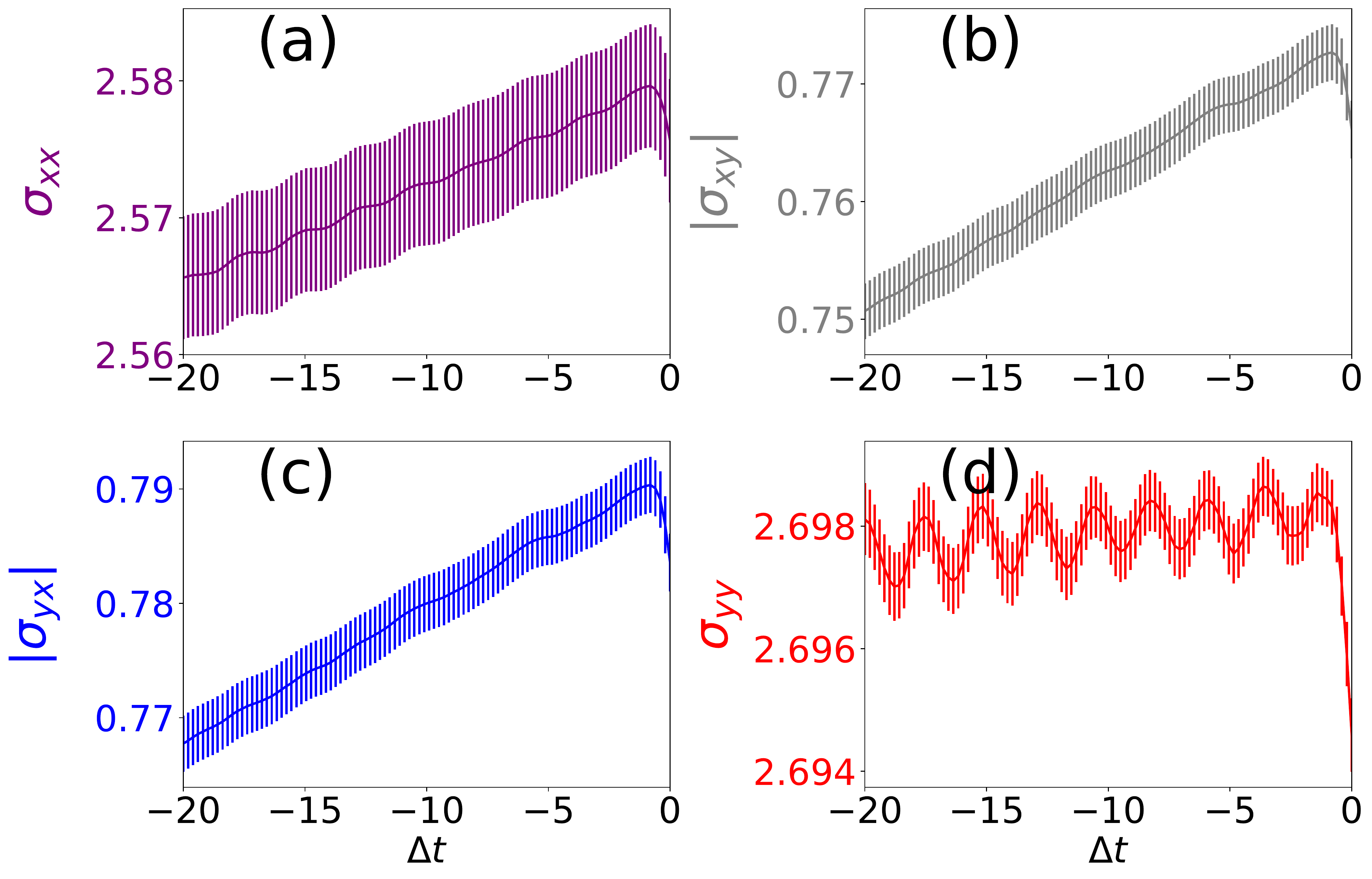}
    \caption{Measures based on the Cauchy tensor averaged over all slip events. (a) $\sigma_{xx}$, (b) modulus of $\sigma_{xy}$, (c) modulus of $\sigma_{yx}$, (d) $\sigma_{yy}$. }
    \label{fig:stresses}
\end{figure}

We also analysed global static measures based on the Cauchy tensor. This tensor is defined as $\sigma_{ij} = \frac{1}{2A}\sum_{c_{k,p}} (F_ir_j + F_jr_i)$, where $A$ is the area of the domain, $r_{i,j}$ are the Cartesian components of the vector pointing from the center of particle $p$ toward
the particle contact $c_k$, and $F_{i,j}$ are the corresponding inter-particle force components. The sum goes over all inter-particle contacts $c_k$ for all particles $p$, excluding the particle–wall contacts. Figure~\ref{fig:stresses} shows the components of the tensor, averaged over all the slip events. We note that from the behavior of these components and the strain results, such as those shown in 
Fig.~\ref{fig:static_pre_slip}(b), one could extract additional information about material response to external 
forcing (such as stiffness tensor). In the 
present context, we just note that the 
behavior of the measures based on the components of the Cauchy tensor shown in Fig.~\ref{fig:stresses} is similar to other global measures.

\begin{figure}[H]
    \centering
    \includegraphics[width = .8\linewidth]{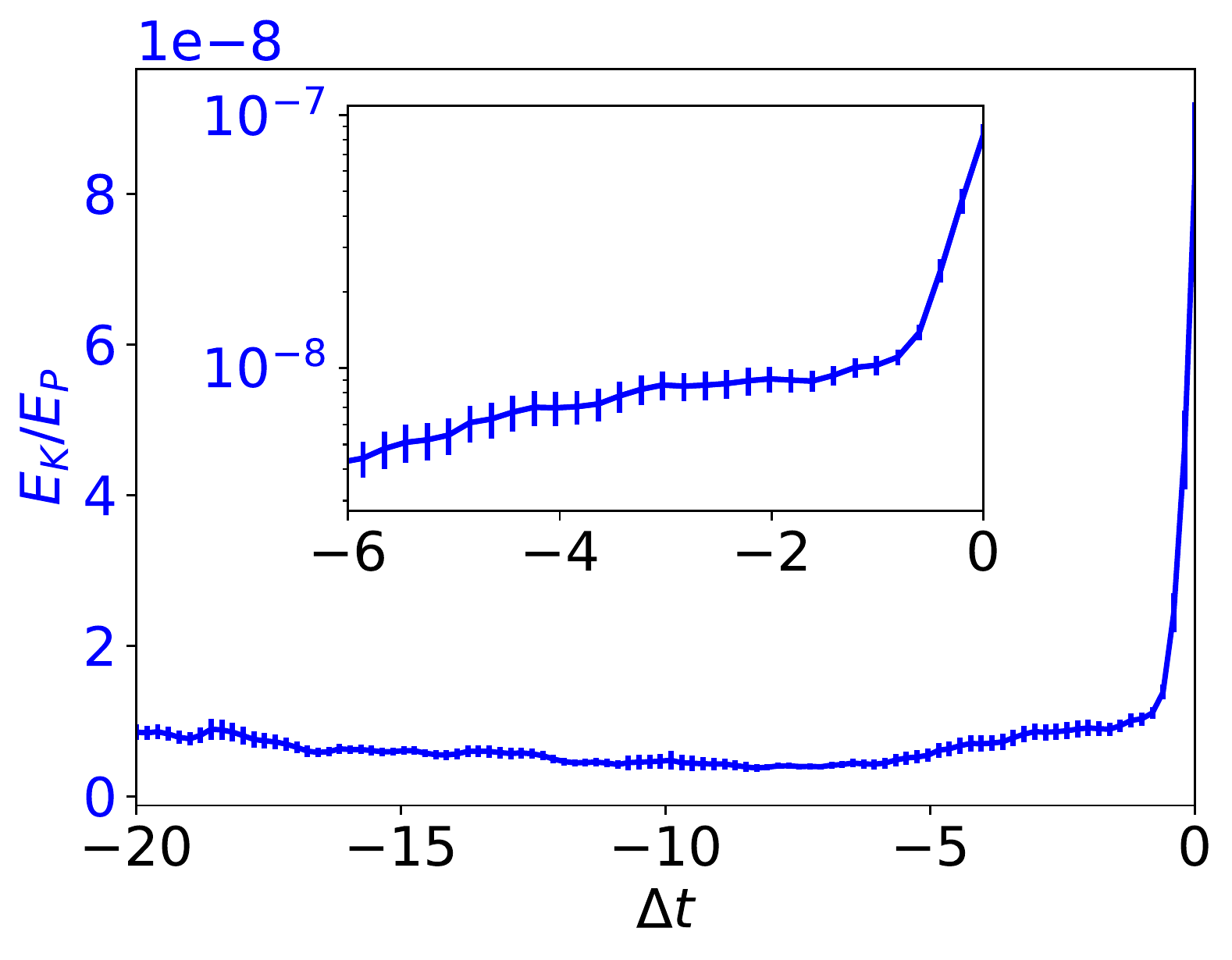}
    \caption{Ratio of Kinetic to potential energy averaged over all slip events. }
    \label{fig:ek/ep}
\end{figure}

Finally, Fig.~\ref{fig:ek/ep} shows the ratio of kinetic and potential energy averaged over all slip events. 
The total kinetic energy $E_k$ = $\sum_{i=1}^{N} m_i{v_i}^2/2$ where $m_i$ and $v_i$ is the mass and the velocity of system particles and the sum is over all the system particles.
We measured total potential energy $E_p$ = $\sum_{C_{i,j}}k_{n}x_{C_{i,j}}^2/2$ where $x_{C_{i,j}}$ is the compression of two particles that are in contact excluding particle-wall contacts. Note that the kinetic energy is obtained by first computing the kinetic energies of the individual particles based on their velocities which are local dynamic quantities. Thus  the ratio of $E_k$ to $E_p$ shows a similar behaviour as the other measures shown in Fig.~\ref{fig:dynamic_pre_slip}.

\section*{Conflicts of interest}
There are no conflicts to declare.

\section*{Acknowledgments}
We acknowledge many useful discussions with late Bob Behringer, Abe Clark, Manuel Carlevaro, Konstantin Mischaikow, Luis Pugnaloni, Joshua Socolar, and Hu Zhang.
Cheng, Basak and Kondic acknowledge support by the ARO Grant No. W911NF1810184.


\balance



\bibliography{granulates.bib}
\bibliographystyle{rsc_ref.bst} 

\end{document}